\documentclass[12pt,a4paper,english,superscriptaddress,aps,nofootinbib]{revtex4}
\usepackage{amsmath,amssymb,graphicx}

\makeatletter
\usepackage{babel}
\usepackage[active]{srcltx}
\usepackage{multirow}
\usepackage{graphicx,color}
\usepackage{changebar}
\usepackage{hyperref}
\usepackage[utf8]{inputenc}
\usepackage[T1]{fontenc}
\usepackage{ae}
\usepackage[caption = false]{subfig}

\usepackage{graphicx}
\usepackage{amsfonts}
\newcommand{\be}{\begin{equation}}
	\newcommand{\ee}{\end{equation}}
\newcommand{\bea}{\begin{eqnarray}}
	\newcommand{\eea}{\end{eqnarray}}

\begin{document}


%
%

\title{Configurational Entropy and braneworlds in $f(T,B)$ gravity}

\author{A. R. P. Moreira\footnote{Email: Allan.moreira@fisica.ufc.br}, \hspace{0.15cm} F. C. E. Lima\footnote{Email: cleiton.estevao@fisica.ufc.br} \hspace{0.15cm} and \hspace{0.15cm} C. A. S. Almeida\footnote{Email: carlos@fisica.ufc.br}}

\address{Universidade Federal do Cear\'a (UFC), Departamento de F\'isica,\\ Campus do Pici, Fortaleza - CE, C.P. 6030, 60455-760 - Brazil.}


\vspace{1cm}
\begin{abstract}
\begin{center}
\vspace{0.25cm}
    \large{Abstract}
\end{center}
The thick brane scenario built on the $f(T,B)$ teleparallel gravity theory was considered for the study of phase transitions, internal structures and new classes of solutions in a model. In this theory, $T$ denotes the torsion scalar, and $B$ is a boundary term. An interesting result was observed when brane splitting occurs, i. e., internal structures in the model arise as a consequence of the appearance of new domain walls in the theory. In fact, this preliminary result influences the profile of the matter field (from kink to multi-kink) so that for appropriate values of the parameters $k_{1,2}$ multiple phase transitions are identified. To perform this analysis, the Differential Configurational Entropy (DCE) that has the ability to predict the existence of phase transitions through critical points was used. Furthermore, the DCE is able to select the most stable solutions since it gives us details about the informational content to the field settings.
\end{abstract}

\maketitle
\newpage




\section{Introduction}

Investigation of the anisotropy of the cosmic microwave background \cite{WMAP,WMAP1}, supernovae \cite{Supernova,Supernova1}, and baryonic acoustic oscillations \cite{SDSS} indicate an accelerated expansion of the universe. This expansion occurs due to the existence of a repulsive gravitational energy, known as dark energy \cite{Sharif}. Currently, there are two lines of investigation that seek to understand the acceleration of cosmic expansion. Several researchers propose the study of exotic energy as an explanation for the problem \cite{Sahni,Sahni1,Padmana,Caldwell,Nojiri,Feng,Kamenshchik}. A new attempt at explaining this question arises when Einstein's Lagrangian is modified. Therefore, modified gravity theories have caught the attention of several researchers \cite{Sharif,Sotiriou,Sotiriou1}.

In 1920, the work of Kaluza \cite{Kaluza} and Klein \cite{Klein} brought the discussion of extra dimensions as a narrative to unify Einstein's general relativity and \cite{Wang} electromagnetism. In 1998, Arkani-Hamed {et. al.} \cite{Arkani} propose a new structure to solve the hierarchy problem that does not depend on supersymmetry or technicolor. In this context, gravitational and gauge interactions become joined at the weak scale. Soon later, L. Randall and R. Sundrum propose a new high-dimensional mechanism to solve the hierarchy problem \cite{Randall&Sundrum,Randall&Sundrum1}.

The braneworlds scenario has been intensively studied in different theories of modified gravity, e. g., the scalar-tensor theory of gravity \cite{Bogdanos,Bogdanos1,Yang,Farakos,Liu,Guo}, the theory of metric $f(R)$ gravity \cite{Parry,Petrov,Petrov1}, and the $f(R)$ Palatini theory \cite{Bazeia,Gu,Olmo,Olmo1}. The theory of  $f(T)$ gravity appears as an alternative to dark energy to explain the acceleration of the universe \cite{Wang,Cai}. These theories have been intensively studied, as seen in Refs. \cite{Behdoodi,Nozari,Moreira2021b,Moreira2021a}.

Recently, new teleparallel gravity models have emerged, namely: the $f(T,T_G)$ gravity with $T_G$ being the Gauss-Bonnet torsion scalar \cite{Kofinas,Kofinas1,Chatto}, $f(T ,\mathcal{T})$ gravity with $\mathcal{T}$ the stress-energy tensor trace \cite{Gomez,Harko}, and the  $f(T,B)$ gravity model with $B$ the term of boundary \cite{Bahamonde,Franco,Rivera}. The $f(T,B)$ model has become an interesting model due to its agreement of theoretical results with experimental data describing the accelerated expansion of the universe \cite{Franco,Rivera,Said}. As a matter of fact, the combination of the scalar torsion and boundary term can allow Lorentz invariant theories \cite{BLi2010,Sotiriou2010,Capozziello2019}. Studies of cosmological perturbations \cite{Bahamonde1}, thermodynamic properties \cite{Pourbagher,Pourbagher1}, and study of gravitational waves \cite{Abedi} have also been developed considering $f(T,B)$ gravity \cite{Bhattacharjee}.

The {\it Mathematical Theory of communication} of C. E. Shannon made it possible to quantify the information and understand how the information can be transferred efficiently from source to the receiver \cite{GleiserSowinki}. Motivated by Shannon's theory, Configuration Information Measures (CIMs) present the proposal to quantify the content and complexity of information contained in a localized physical structure. These structures arise in topological field theories, as can be seen in Refs. \cite{Ranada, Dva, LPA}. Formulated in the reciprocal space, Configurational Entropy (CE) and Configurational Complexity (CC), and their differential variants, DCE and DCC, are the bases of CIMs. CE plays an important role in theory, i. e., quantifies the number of bits needed to build a stable field configuration out of wave modes \cite{GleiserSowinki,GS,Correa2015a}.

In 2012, Gleiser and Stamatopoulos (GS) \cite{Gleiser2011} show that CE brings information about some parameters of a given model for which the energy density is located. In Ref. \cite{Gleiser2011} it was shown that the higher the energy that approximates the actual solution, the higher its relative CE, which is defined as the absolute difference between the actual function and the test function of the CE \cite{Correa2015a}. CE applications have been shown to be a promising path for the physical understanding of some systems, e. g., the non-equilibrium dynamics of spontaneous symmetry breaking \cite{G1}, the stability bound for compact objects \cite{G2}, and for investigate the emergence of objects located during the inflationary preheating \cite{G3}.

In braneworld scenario, DCE can be used to measure the degree of informational organization in the structures of the system. In fact, DCE plays an important role to decide the most suitable intrinsic parameters of the models \cite{Correa2015c,Correa2016a,Correa2016b, Correa2016p}. The approach of the calculate of DCE can be applied to study the formation of internal structures in the brane \cite{Cruz2017,Cruz2018}. Additionally, DCE was studied in braneworld scenarios in a context of modified gravity of the type $f(R)$ \cite{Correa2015a}, and $f(R,\mathcal{T})$ \cite{Correa2015b}. Inspired by these results, our purpose here is make a more accurate analysis of the conditions that lead to the phase transition in a braneworld scenario in a modified teleparallel $f(T,B)$ gravity. For this the DCE is calculated in our model.

Our work is organized as follows: In Sec. II, the fundamental concepts of $f(T, B)$ gravity in a braneworld scenario is presented. In addition, the  energy densities of the brane and the matter field solutions of the models $f_1(T,B)=T+k_1B^{n_1}$, and $f_2(T,B)=T +k_2(-T+B)^{n_2}$ were investigated. In Sec. III, a brief review of CE and DCE concepts is performed. Then, we study the DCE in the mentioned models. Finally, in Sec. IV we discuss our findings.

\section{$f(T,B)$ brane models}
\label{sec1}

Motivated by recent works on modified teleparallel $f(T,B)$ gravity \cite{Moreira2021a}, in this section we present the fundamental concepts of the $f(T,B)$ gravity  in a braneworld scenario.

\subsection{$f(T,B)$ gravity}

In teleparallel gravity, the dynamic variable for $f(T,B)$ gravity are the \textit{vielbein}, which are related to metric as follows,
\begin{eqnarray}
g_{MN}=\eta_{ab}h^a\ _M h^b\ _N,
\end{eqnarray}
where the metric signature is $\eta_{ab}=diag(-1,1,1,1,1)$, the uppercase Latin letters $M,N=0, 1, 2, 3, 4$ represent the indices of the bulk coordinates and the lowercase Latin letters $a,b=0, 1, 2, 3, 4$ represent the indices of the tangent space coordinates.

The interesting thing about teleparallel gravity is that it assumes an non-curvature connection, which is known as  Weitzenb\"{o}ck connection $\widetilde{\Gamma}^P\ _{NM}=h_a\ ^P\partial_M h^a\ _N$ \cite{Aldrovandi}. In this context, the torsion is not null, and is 
represented in terms of the Weitzenb\"{o}ck connection as \cite{Aldrovandi}
\begin{eqnarray}
T^{P}\  _{MN}= \widetilde{\Gamma}^P\ _{NM}-\widetilde{\Gamma}^P\ _{MN}.
\end{eqnarray}

Through torsion, the contorsion tensor \cite{Aldrovandi} is define as 
\begin{eqnarray}
K^P\ _{NM}=\frac{1}{2}\Big( T_N\ ^P\ _M +T_M\ ^P\ _N - T^P\ _{NM}\Big).
\end{eqnarray}
On the other hand, the dual torsion tensor \cite{Aldrovandi} is
\begin{eqnarray}
S_{P}\ ^{MN}=\frac{1}{2}\Big( K^{MN}\ _{P}-\delta^N_P T^{QM}\ _Q+\delta^M_P T^{QN}\ _Q\Big),
\end{eqnarray} 
and the torsion scalar $T=T_{PMN}S^{PMN}$. The curvature scalar and the torsion scalar are related as follows,
\begin{eqnarray}\label{1212}
R=-T+B,
\end{eqnarray}
where
\begin{eqnarray}
B\equiv -2\nabla^{M}T^{N}\ _{MN}=\frac{2}{h}\partial_M(h T^M)
\end{eqnarray}
is the boundary term, being that $T_M=T^N\ _{MN}$.

The gravitational Lagrangian is written in terms of $T$ and $B$, i. e., $\mathcal{L}=-hf(T, B)/4\kappa_g $, where $h=\sqrt{g} $, and $\kappa_g=4\pi G/c^4$ is the gravitational constant \cite{Abedi2017}. So, the gravitational action can be written as 
\begin{eqnarray}\label{55.5}
\mathcal{S}=-\frac{1}{4\kappa_g}\int h \Big[f(T,B)+4\kappa_g \mathcal{L}_m\Big]d^5x,
\end{eqnarray}
where $\mathcal{L}_m$ is the matter Lagrangian. 

By varying the action (\ref{55.5}) in terms of the dynamic variable (\textit{vielbein}), we obtain the field equations, namely,
\begin{eqnarray}\label{3.36}
\frac{1}{h}f_T\Big[\partial_Q(h S_N\ ^{MQ})-h\widetilde{\Gamma}^R\ _{SN}S_R\ ^{MS}\Big]+\frac{1}{4}\Big[f-Bf_B\Big]\delta_N^M& &\nonumber\\+\Big[(\partial_Qf_T)+(\partial_Qf_B) \Big]S_N\ ^{MQ} +\frac{1}{2}\Big[\nabla^M\nabla_N f_B-\delta^M_N\Box f_B\Big]&=&\kappa_g\mathcal{T}_N\ ^M,
\end{eqnarray}
where $f\equiv f(T,B)$, $f_T\equiv\partial f(T,B)/\partial T$, $f_{B}\equiv\partial f(T,B)/\partial B$, $\Box\equiv\nabla^M\nabla_M$  and $\mathcal{T}_N\ ^M$ is the stress-energy tensor, that arises from the Lagrangian variation of matter in terms of \textit{vielbein}.

\subsection{$f(T,B)$ braneworld}

To analyze the braneworld scenario in a modified teleparallel $f(T,B)$ gravity, the following metric ansatz is considered  \cite{Yang2012},
\begin{equation}\label{45.a}
ds^2=e^{2A(y)}\eta_{\mu\nu}dx^\mu dx^\nu+dy^2,
\end{equation}
where $e^{A(y)}$ is the warp factor and $\eta_{\mu\nu}=(-1, 1, 1, 1)$ is the Minkowski metric. 

For the metric (\ref{45.a}), we choose \textit{vielbein} in the form $h^a\ _M = diag (e^A,e^A,e^A,e^A, 1)$, which represents a good choice, since the equations of the gravitational field do not add any additional restrictions to the functions $f(T,B)$ and neither to the torsion scalar and the boundary term \cite{Moreira2021a,Moreira2021b}. Namely,
\begin{eqnarray}
 T=-12A'^2\ \ \ \text{and} \ \ \ B=-8(A''+4A'^2),
\end{eqnarray}
where the prime $ (\ '\ ) $ represents differentiation from $y$. Since the curvature scalar can be written in terms of the torsion scalar and the boundary term (\ref{1212}), we have 
\begin{eqnarray}
R=-4(5A'^2+2A'').
\end{eqnarray}

Finally, we add a matter Lagrangian with a real scalar field $\phi\equiv \phi(y)$, given in the form
\begin{equation}
\mathcal{L}_m=\frac{1}{2}\partial^M\phi\partial_M\phi+V(\phi).
\end{equation}
The energy density is obtained as follows,
\begin{equation}\label{3333}
\rho(y)=-e^{2A}\mathcal{L}_m.
\end{equation}

With Eq. (\ref{3.36}), the gravitational field equations can be written as,
\begin{align}
&\phi''+4A'\phi'=\frac{\partial V}{\partial \phi},\label{scalarfieldeom}\\
&\frac{1}{4}\Big[f+8(A''+4A'^2)f_B\Big]+6A'^2f_T=\kappa_g\Big(\frac{1}{2}\phi'^2-V \Big),\label{e.1}\\ \nonumber
&12\Big[A'(A'''+8A'A'')(f_{BB}+f_{TB})+3A''A'^2(f_{TT}+f_{BT})\Big]-\frac{1}{2}(A''+4A'^2)\times\\
&(4f_{B}+3f_{T})-\frac{1}{4}f=\kappa_g\Big(\frac{1}{2}\phi'^2+V \Big).\label{e.2}
\end{align}

Although Eqs. (\ref{scalarfieldeom}), (\ref{e.1}) and (\ref{e.2}) look simple, it is difficult to find an analytical solution for the most general cases. For simplicity, let us propose an ansatz warp factor of the form \cite{Gremm1999},
\begin{eqnarray}
\label{coreA}
e^{2A(y)}=\cosh^{-2p}(\lambda y),
\end{eqnarray}
where $p$ and $\lambda$ are the parameters that determine the width and width of the source, respectively.

Now we propose two models of $f(T,B)$, i. e., $f_1(T,B)=T+k_1B^{n_1}$, that is a model already analyzed in the braneworld scenario \cite{Moreira2021a,Moreira2021b}, and $f_2(T,B)=T+k_2(-T+B)^{n_2}$ which gives a very interesting model. This choice of $f(T,B)$ leads to results very similar to that obtained in Ref. \cite{Correa2015b}, where is considered a $f(T,R)$ gravity model with $f(T,R)=T+kR^{n}$. The $k_{1,2}$ and $n_{1,2}$ parameters control the modification of the usual teleparallel theory. To simplify our analysis, the gravitational constant $\kappa_g=1$ is considered.

\subsubsection{$f_1(T,B)=T+k_1B^{n_1}$}

As shown in Refs. \cite{Moreira2021a,Moreira2021b}, for $f_1(T, B)$ the equations of the metric components and the scalar field are 
\begin{eqnarray}\label{q.5} \nonumber
\phi'^2(y)=&\frac{3}{2}p\lambda^2\mathrm{sech}^2(\lambda y)-\frac{\alpha^{n_1}}{(2\beta-1)^2}\Big[8^{n_1-1}k_1n_1(n_1-1)(1+4p)\sinh^2(\lambda y)\Big],\\
\end{eqnarray}
\begin{eqnarray}\label{q.55}
V(\phi(y))=&\frac{3}{4}\alpha+2^{3n_1-4}k_1(n_1-1)(p\lambda^2)^2\alpha^{n_1-2}\Big\{4 \mathrm{sech}^4(\lambda y)+64p^2\tanh ^4(\lambda y)\nonumber\\ &-[3n_1+4(8+3n_1)]\mathrm{sech}^2(\lambda y)\tanh ^2(\lambda y)\Big\},
\end{eqnarray}
where $\alpha\equiv p \lambda^2[\mathrm{sech}^2(\lambda y)-4p\tanh^2(\lambda y)]$ and $\beta\equiv1+p-p\cosh(2\lambda y)$.
With equation (\ref{3333}), we can find the energy density for $f_1(T,B)$, which is
\begin{eqnarray}
\rho(y)&=&\Bigg\{2^{3n_1-2}\alpha(n_1-1)k_1 -\frac{8^{n_1-1}}{(2\beta-1)^2}\Big[\alpha(n_1-1)3n_1k_1(1+4p)\sinh^2(\lambda y)\Big]\nonumber\\
& &-3\Big[(p\lambda)^2-\frac{1}{2}(1+2p)p\lambda^2\mathrm{sech}^2(\lambda y)\Big]\Bigg\}\cosh^{-2p}(\lambda y).
\end{eqnarray}

\begin{figure}[ht!]
\begin{center}
\begin{tabular}{ccc}
\includegraphics[height=5cm]{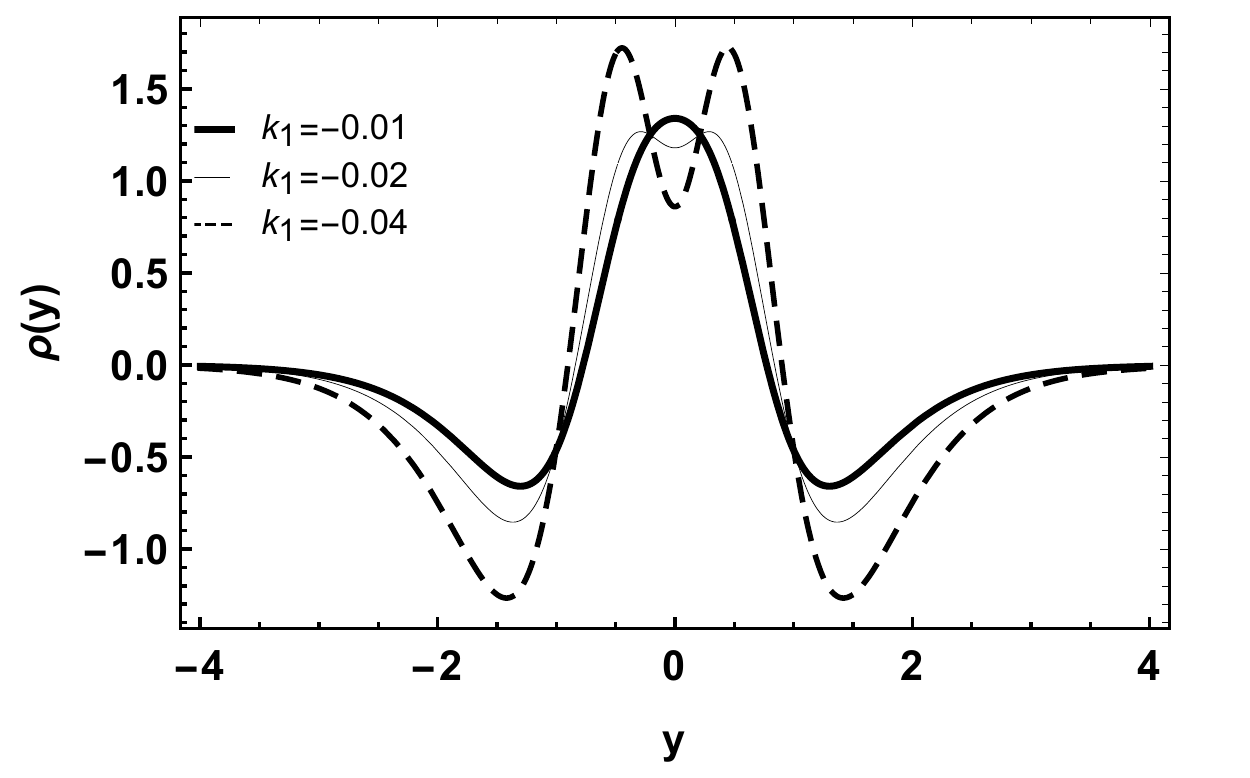}
\includegraphics[height=5cm]{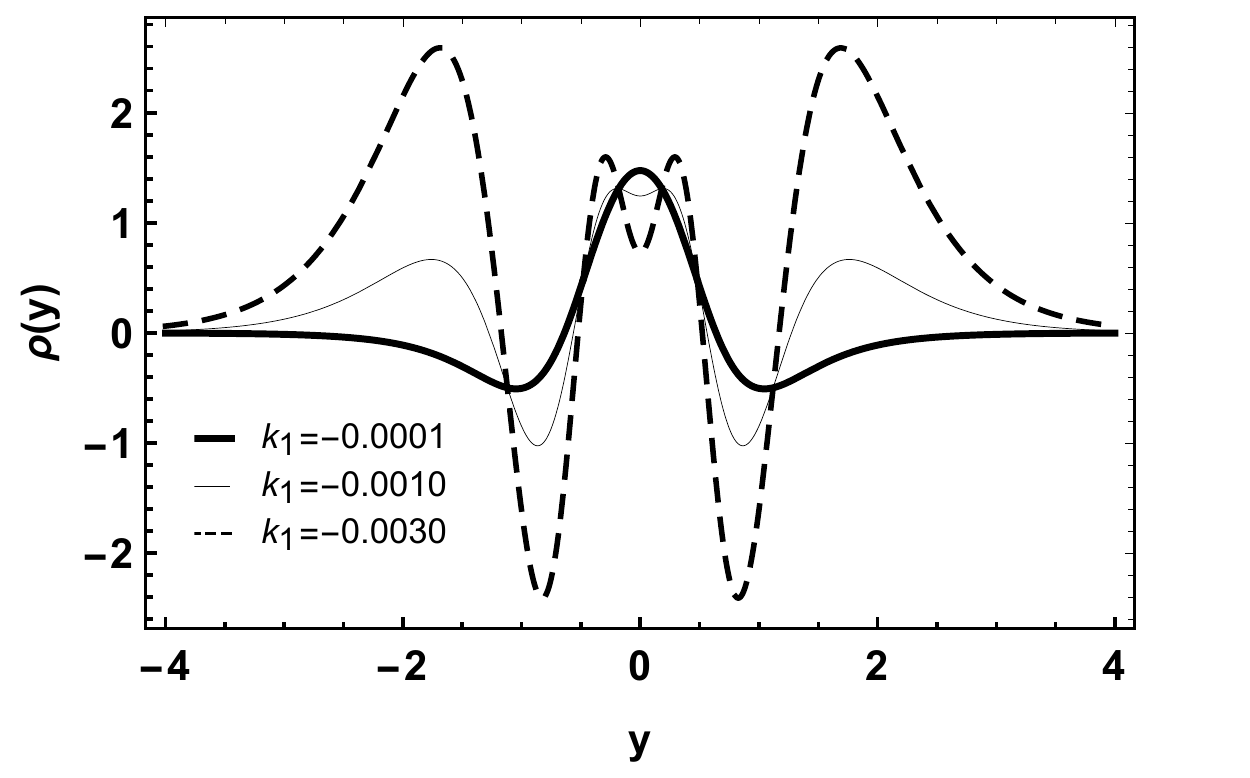}\\ 
\hspace{1cm} (a) \hspace{7.5cm} (b)
\end{tabular}
\end{center}
\vspace{-0.7cm}
\caption{The energy density for $p=\lambda=1$. (a) $n_1=2$ . (b) $n_1=3$.
\label{fig1}}
\end{figure}

The behavior of the energy density of the brane is shown in Fig. \ref{fig1}. For the case $n_1=2$ (Fig.\ref{fig1}$a$), varying the value of the parameter $k_1$ is observed the emergence of internal structures in the brane. Further, for $n_1=3$ (Fig.\ref{fig1}$b$), varying the parameter $k_1$ the internal structures of the model are intensified (this is seen with the increase amount of maximal points). In fact, the appearance of such structures is related to the division of the brane \cite{Moreira2021b,Moreira2021a}.

\begin{figure}[ht!]
\begin{center}
\begin{tabular}{ccc}
\includegraphics[height=3.5cm]{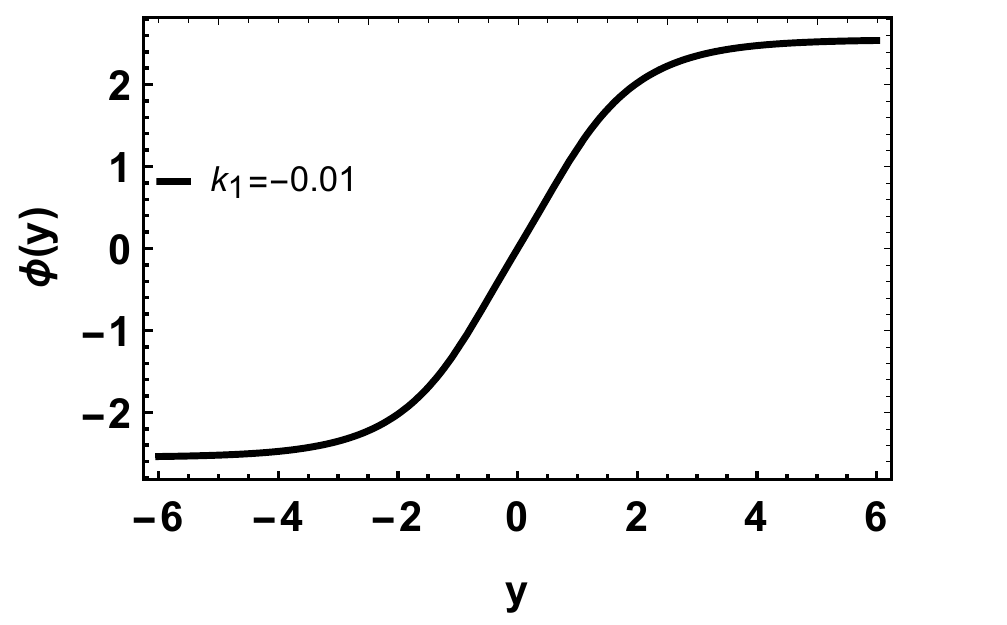}
\includegraphics[height=3.5cm]{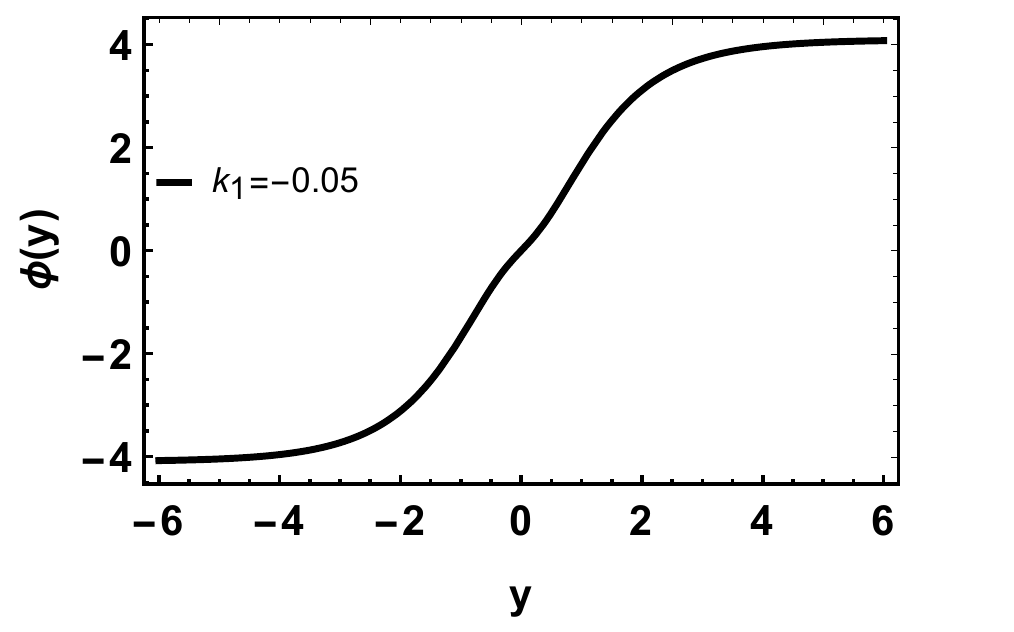}
\includegraphics[height=3.5cm]{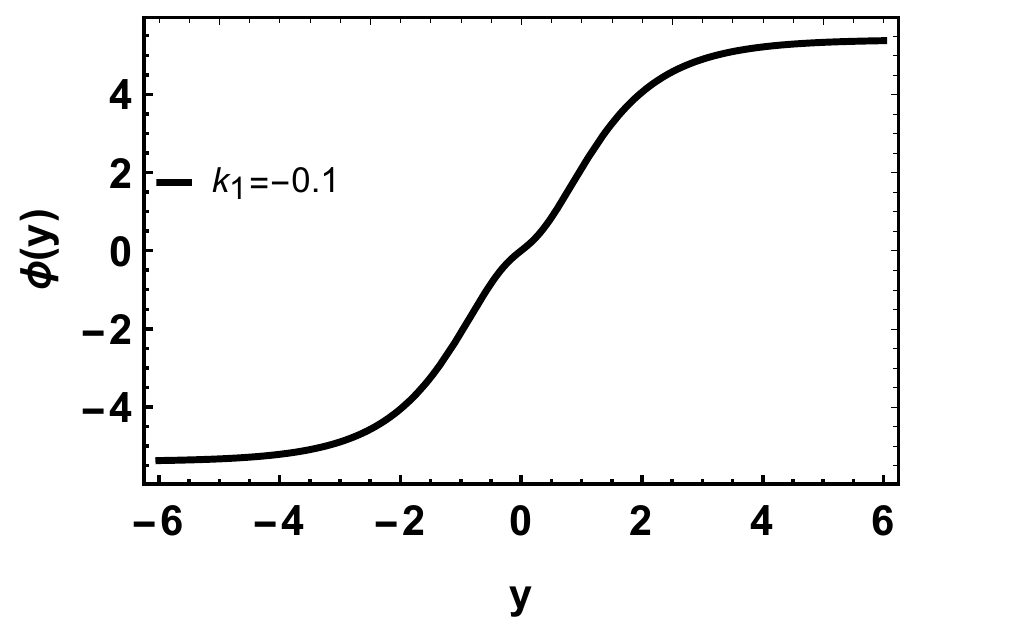}
\end{tabular}
\end{center}
\vspace{-0.7cm}
\caption{
The scalar field solution for $n_1=2$ with $p=\lambda=1$.
\label{fig2}}
\end{figure}

\begin{figure}[ht!]
\begin{center}
\begin{tabular}{ccc}
\includegraphics[height=3.5cm]{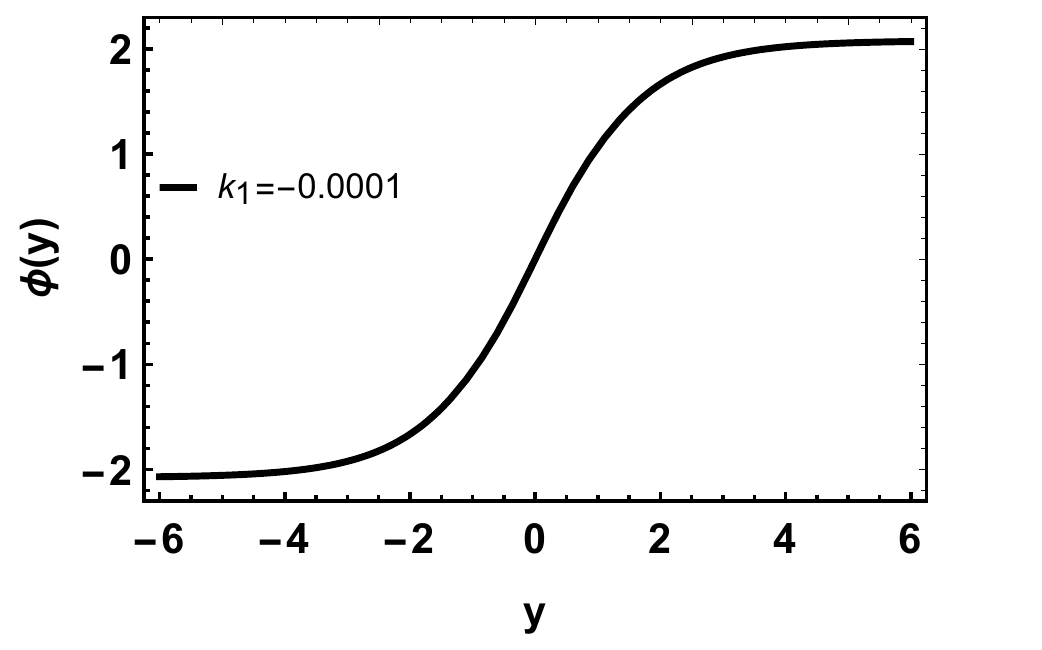}
\includegraphics[height=3.5cm]{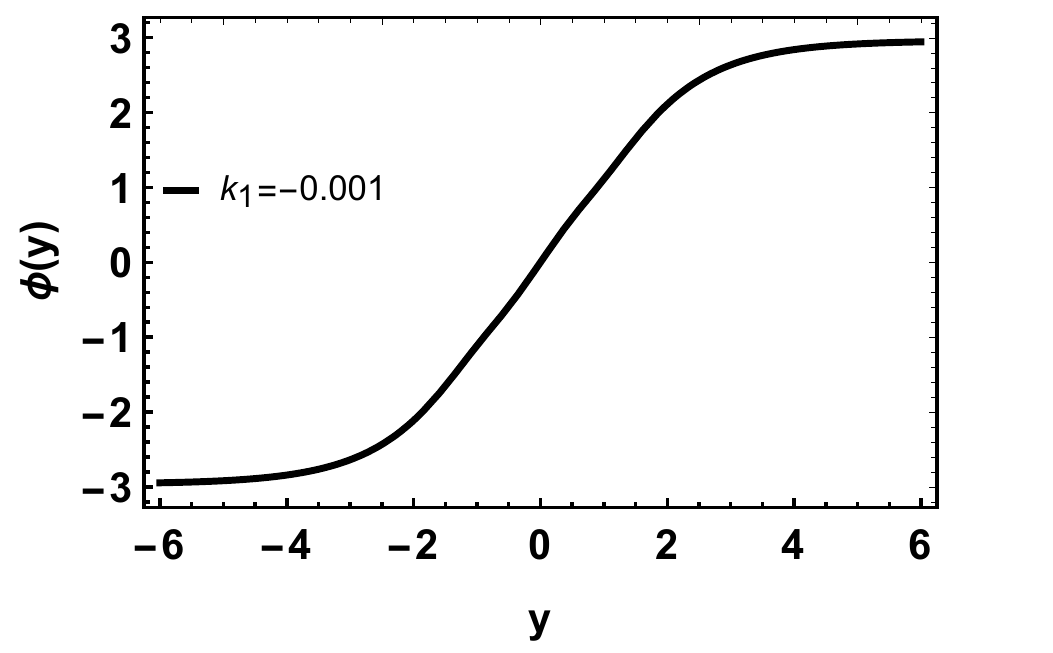}
\includegraphics[height=3.5cm]{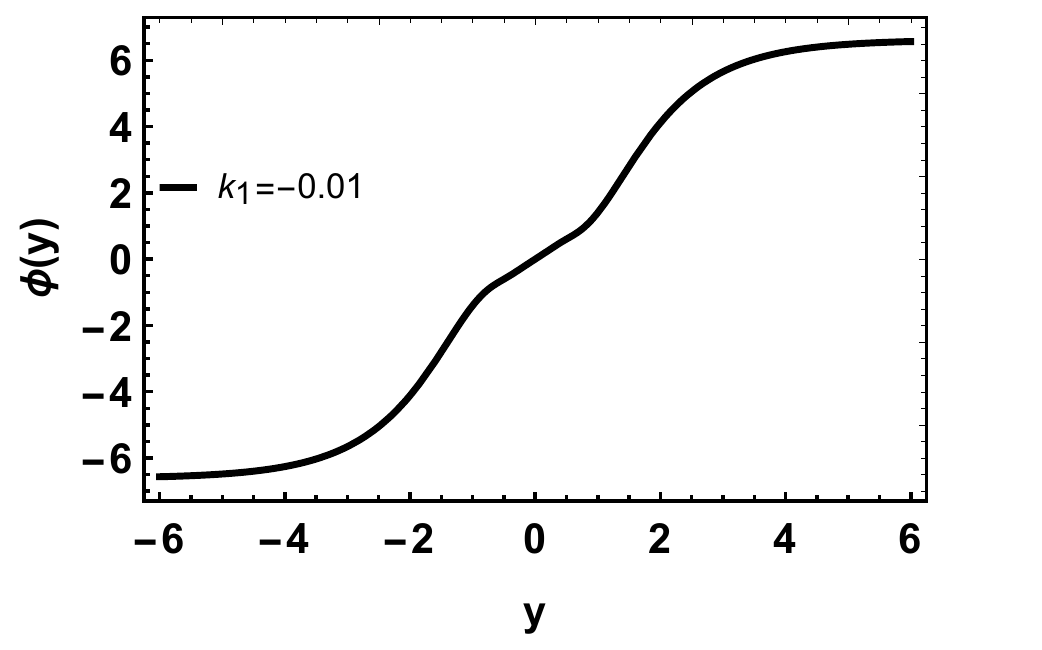}
\end{tabular}
\end{center}
\vspace{-0.7cm}
\caption{The scalar field solution for $n_1=3$ with $p=\lambda=1$.
\label{fig3}}
\end{figure}

Through Eq. (\ref{q.5}) we obtain the scalar field. As the general expression for the scalar field is difficult to obtain, the numerical solutions for two particular cases $n_1=2$ and $n_1=3$ are investigated. Figs. \ref{fig2} and \ref{fig3} describe the behavior of the scalar field (Fig. \ref{fig2} for $n_1=2$ and Fig. \ref{fig3} for $n_1=3$). It seems to us that the model supports multiple phase transitions transforming kink-like solutions into more complex configurations as the brane splitting occurs.

A particularly interesting result is the existence of branes with internal structure for certain values of $k_1$. As seen in Ref. \cite{D}, the existence of such structures in low-dimensional models coupled by gauge fields lead to the existence of structures with superconducting properties. In the study of cosmic strings the internal structures are related to the presence of superconducting strings (see for example \cite{Hind,Hart,Cao}). At this point, let us to give a possible interpretation of the phenomenon in our model. Note that for the value of $k_1=-0.01$ and $n_1=2$ the topological scalar field configuration have a kink-like behavior. In this case, the expected behavior for energy density when $y\to 0$ is $\rho(y)\approx$ sech$^2(y)$. However, in the case that $n_1$ as well as the parameter $k_1$ decrease, the appearance of internal structures on the brane it becomes evident. This is seen as a direct consequence of a new domain wall that starts to form in the vicinity of $ y=0$. The appearance of these new domain walls makes the topological configuration of the scalar field (kink-like) to have a profile similar to a double-kink or even multi-kink configuration. Here it is interesting to mention that similar behavior occurs for other values of $n_i$, so this discussion can be generalized to all cases showed in this work. We return to this discussion in the next section when we introduce the DCE formalism to identify the location of domain walls and model phase transitions.

\subsubsection{$f_2(T,B)=T+k_2(-T+B)^{n_2}$}

Using Eqs. (\ref{e.1}) and (\ref{e.2}), we obtain the equations that are related to the metric components and the scalar field as

\begin{eqnarray}\label{q.6}
\phi'^2(y)=\frac{3}{2}\Big\{1-4^{n_2-1}k_2n_2[(2+5p)\mathrm{sech}^2(\lambda y)-5p]^{n_2-1}\Big\}p\lambda^2\mathrm{sech}^2(\lambda y),
\end{eqnarray}
\begin{eqnarray}\label{q.606}
V(\phi(y))&=&\frac{1}{8}\Bigg\{6p\lambda^2(\beta-1)\mathrm{sech}^2(\lambda y)-2{n_2}k_2(5\beta-1)^{n_2-1}[p\lambda^2\mathrm{sech}^2(\lambda y)]^{n_2}\nonumber\\& &\times\Big[8-5n_2+4p(2n_2-5)\sinh^2(\lambda y)\Big]\Bigg\},
\end{eqnarray}
remembering that $\beta\equiv1+p-p\cosh(2\lambda y)$. The energy density is obtained for $f_2(T,B)$ using Eq.(\ref{3333}), which here has the form
\begin{eqnarray}
\rho(y)&=&\frac{1}{4}\Big\{(4p\lambda^2)^{n_2}k_{2}[(2+5p)\mathrm{sech}^2(\lambda y)-5p]^{n_2}\Big[\frac{5\beta-1-n_2(2\beta-1)}{1-5\beta}\Big]\nonumber\\& &+6p\lambda^2\beta\sinh^2(\lambda y)\Big\}\cosh^{-2p}(\lambda y),
\end{eqnarray}

\begin{figure}[ht!]
\begin{center}
\begin{tabular}{ccc}
\includegraphics[height=5cm]{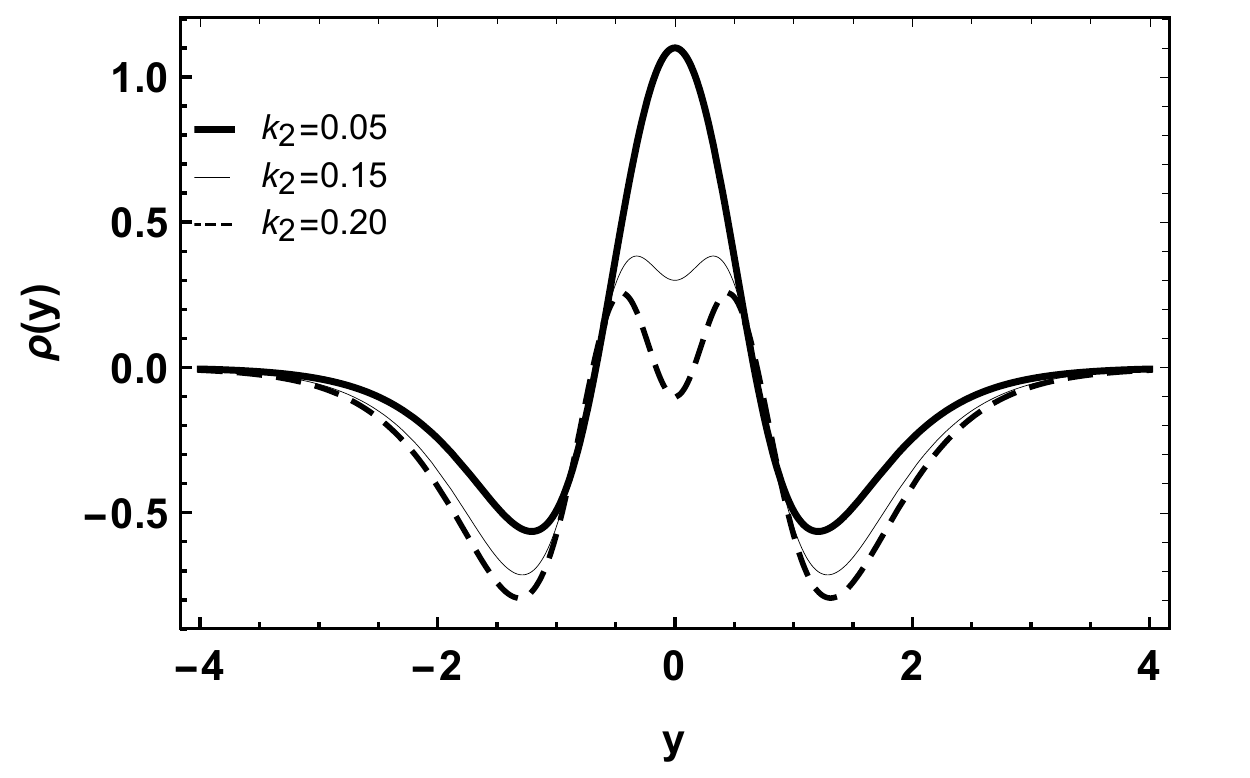}
\includegraphics[height=5cm]{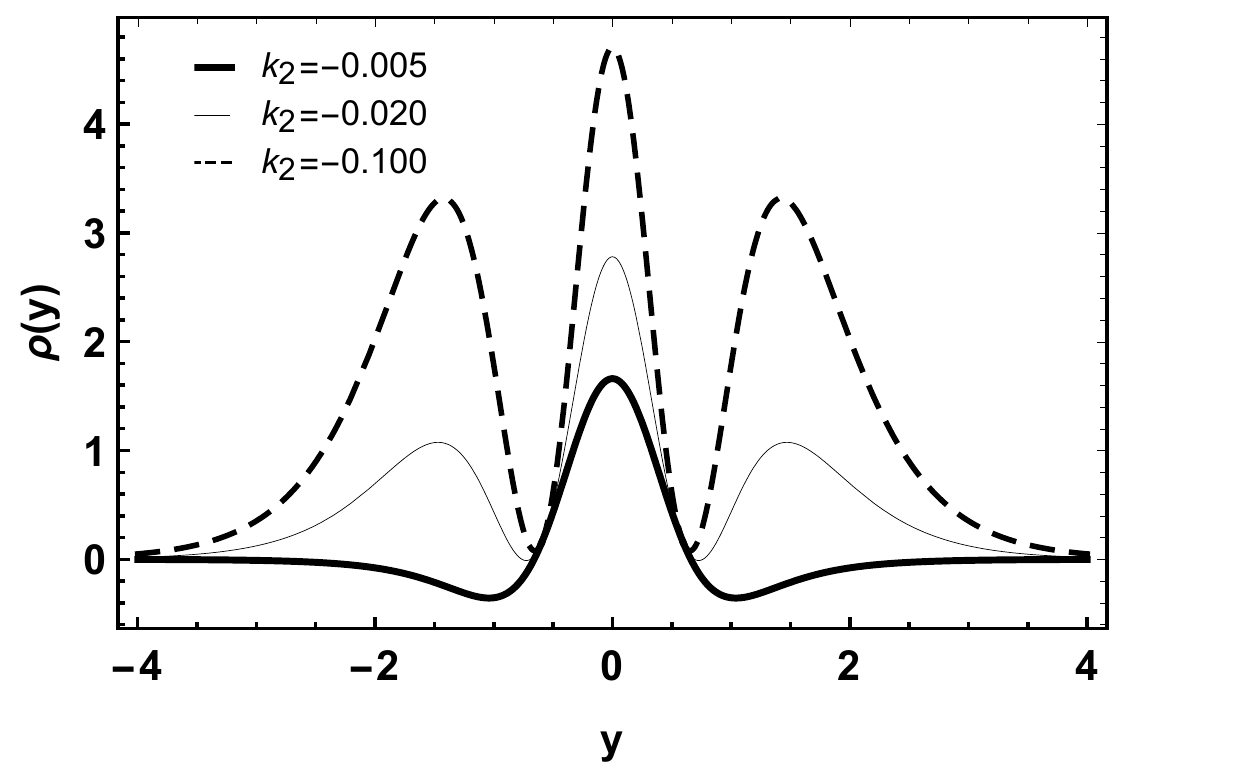}\\ 
(a) \hspace{8 cm}(b)
\end{tabular}
\end{center}
\vspace{-0.7cm}
\caption{The energy density for $p=\lambda=1$. (a) $n_2=2$ . (b) $n_2=3$.
\label{fig4}}
\end{figure}

The energy densities of the brane are shown in Fig. \ref{fig4}. For $n_2=2$ (Fig. \ref{fig4}$a$), varying the value of the parameter $ k_2$, arise the internal structures in the model, i. e., two symmetric (maximum) critical points appear near the origin. Meanwhile, the wells around the structures increase in depth. For $n_2=3$ (Fig. \ref{fig4}$b$), varying the value of the parameter $k_2$, we have the appearance of structures with three peaks. This energy density profile shows the existence of localized structures. 

Now, let us draw attention to the region around $y=0$ of the  energy density. In this region, it is evident for this case that there is multiple phase transition making the energy density more intense and located around the origin. In fact, analyzing the profile of the scalar field is observed that this is a consequence of the emergence of settings similar to multi-kink. Here it is interesting to mention that these results are related to brane splitting.

\begin{figure}[ht!]
\begin{center}
\begin{tabular}{ccc}
\includegraphics[height=3.5cm]{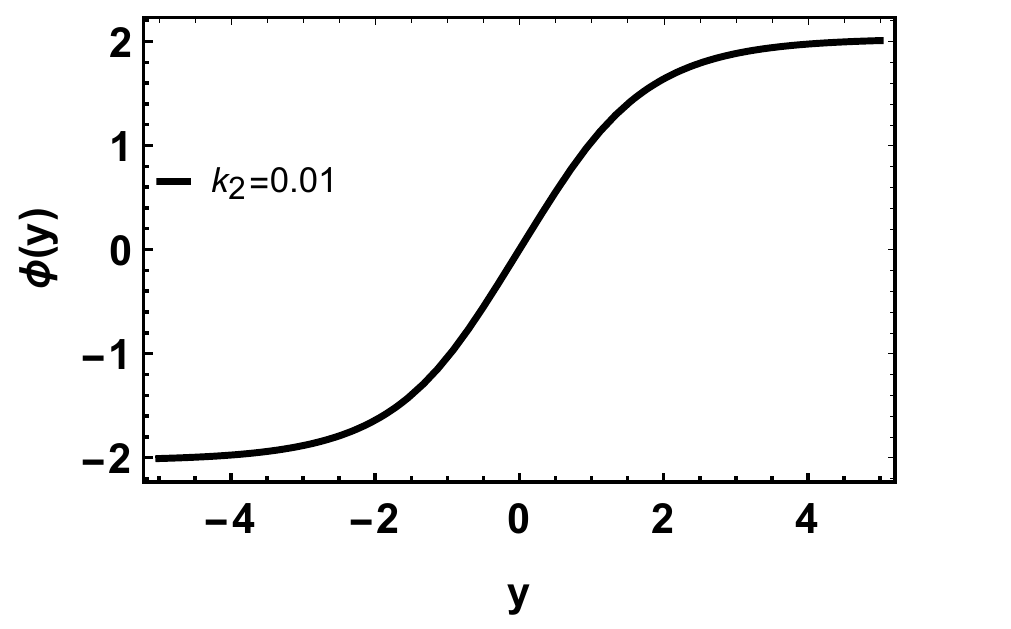}
\includegraphics[height=3.5cm]{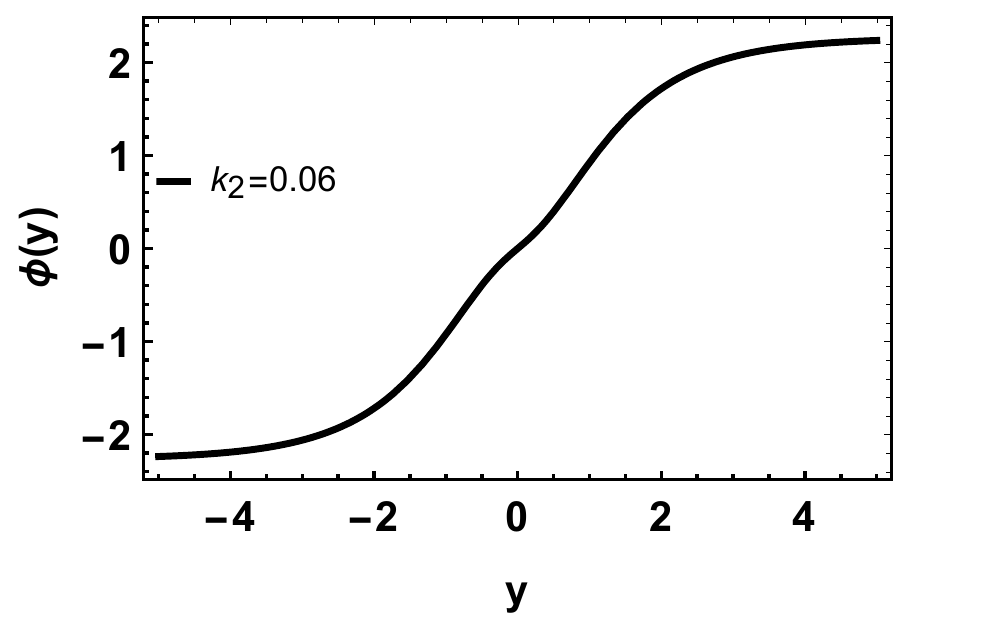}
\includegraphics[height=3.5cm]{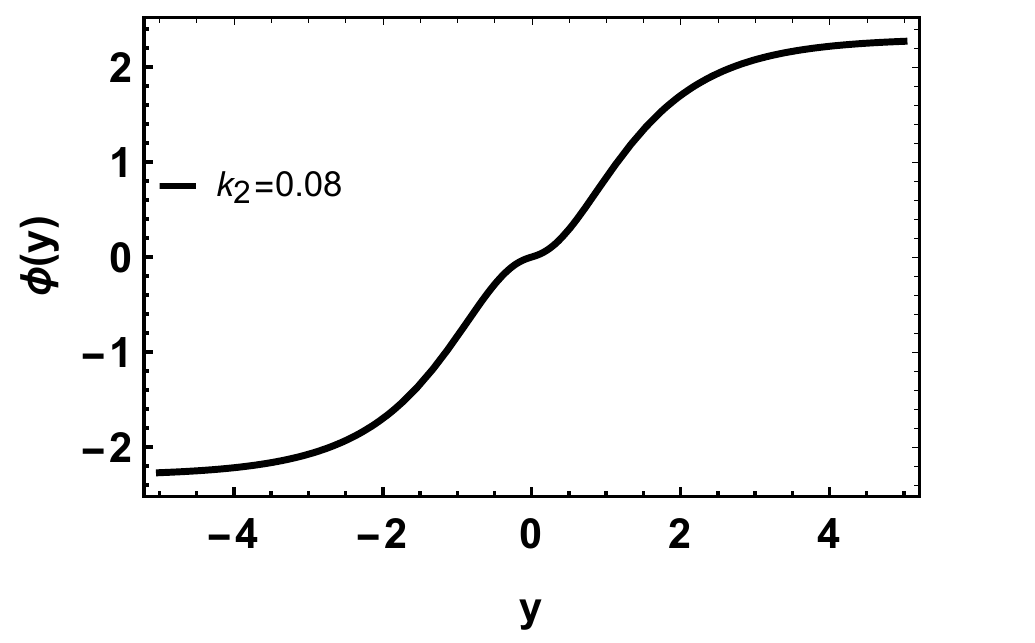}
\end{tabular}
\end{center}
\vspace{-0.7cm}
\caption{The scalar field solution for $n_2=2$ with $p=\lambda=1$.
\label{fig5}}
\end{figure}

\begin{figure}[ht!]
\begin{center}
\begin{tabular}{ccc}
\includegraphics[height=3.5cm]{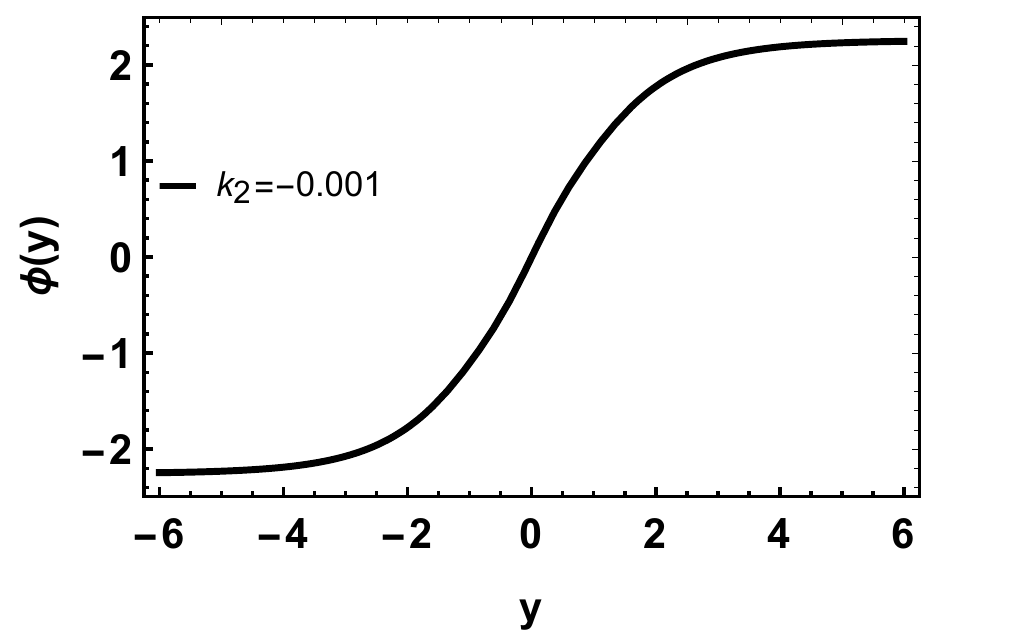}
\includegraphics[height=3.5cm]{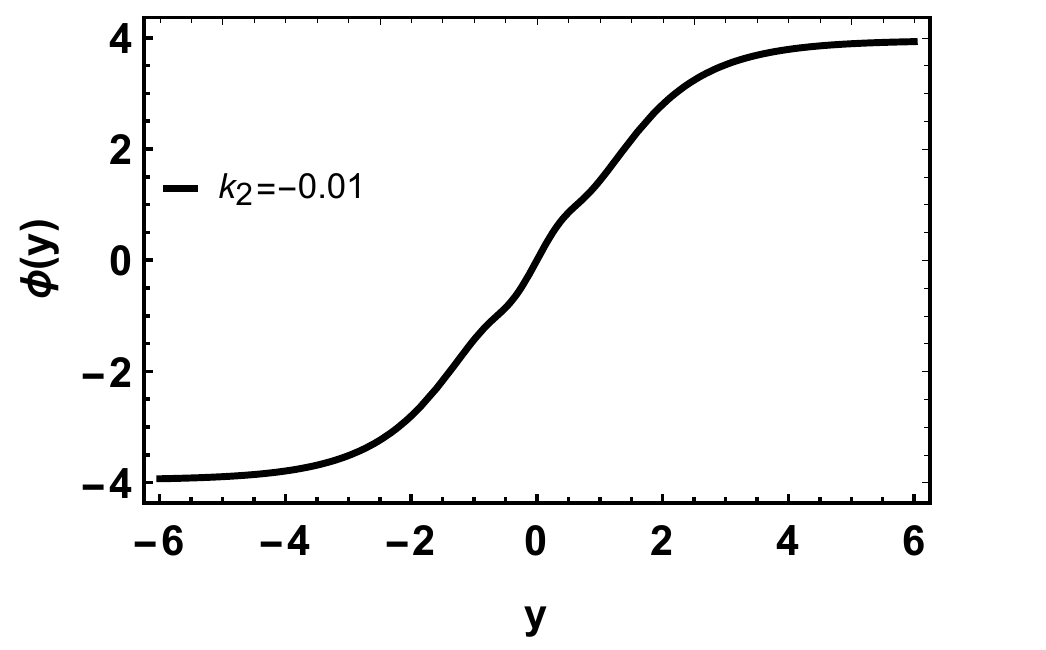}
\includegraphics[height=3.5cm]{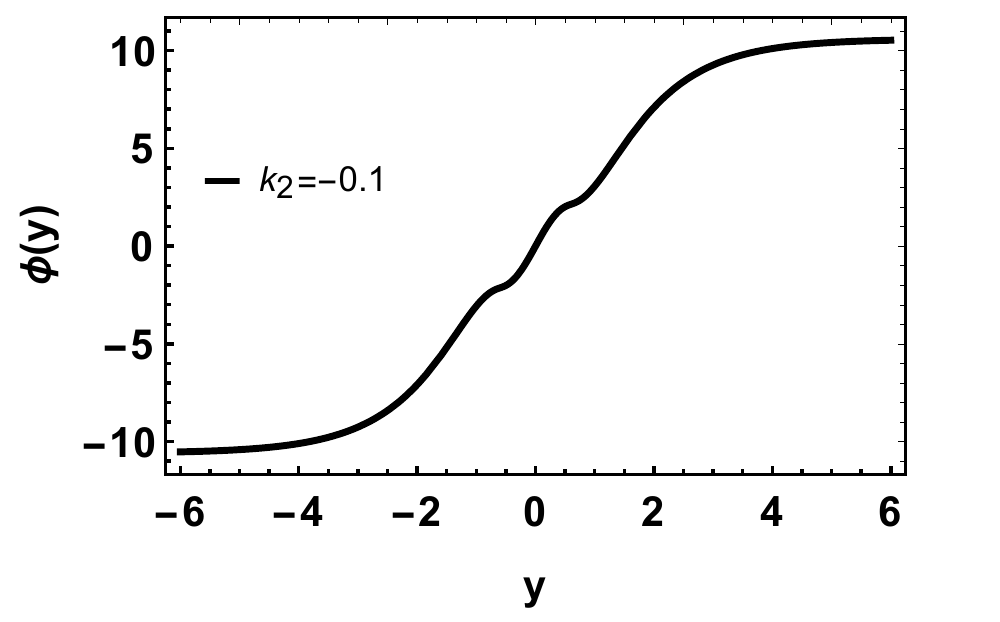}
\end{tabular}
\end{center}
\vspace{-0.7cm}
\caption{The scalar field solution for $n_2=3$ with $p=\lambda=1$.
\label{fig6}}
\end{figure}

Using Eq. (\ref{q.6}) the scalar field is obtained. For simplicity, we solve numerically the equation since the general expression for the scalar field is difficult to obtain. The behavior of the solutions of the field for two particular cases, i. e.,  $n_2=2$, and $n_2=3$ are presented respectively in Figs. \ref{fig5} and \ref{fig6}.  In fact, it is evident that the scalar field assumes a behavior kink or multi-kink type (in some case). It seems to us that for some values of $k_2$ we have the appearance of a new domain wall, and consequently, multiple phase transitions when the brane is splitting. This will be detailed in the next section.

\section{Configurational Entropy in brane Models}
\label{sec2}

In 2012, Gleiser and Stamatopoulos (GS) \cite{GS} suggest a representation of the detailed measure of the complexity of fields , i. e., the so-called configurational entropy (CE). As discussed in Ref. \cite{GleiserSowinki} the configurational information measures (CIMs) arises from the desire to quantify the informational content and complexity contained in the form of localized structures that arise in field theories. Constructed in reciprocal space, configurational entropy and its differential variant, e. g., differential configurational entropy (DCE), is a concept that is at the core of CIMs. In this section, a brief discussion of the property of DCE measurements in a braneworld scenario is presented. Using the concept of DCE, the possible phase transitions that the model supports are investigated.

To calculate the DCE, let us define the Fourier transform as discussed in Refs. \cite{Gleiser2011,G2}, namely,
\begin{eqnarray}\label{4444}
\mathcal{F}[\omega]=\frac{1}{\sqrt{2\pi}}\int e^{i\omega y}\rho(y) dy.
\end{eqnarray}
In the braneworld $\mathcal{L}_m$ is the Lagrangian density of matter and $e^{2A}$ is the warp factor. As defined in Eq. (\ref{3333}), the energy density is $\rho(y)=-e^{2A}\mathcal{L}_m$ \cite{Correa2015c,Correa2016a}, so Eq. (\ref{4444}) takes the form
\begin{eqnarray}
\mathcal{F}[\omega]=-\frac{1}{\sqrt{2\pi}}\int e^{2A(y)+i\omega y}\mathcal{L}_m dy.
\end{eqnarray}

We define now the modal fraction $f(\omega)$, which measures the relative weight of each mode $\omega$ as \cite{Gleiser2011, G2,Correa2015a,Correa2015b,Correa2015c,Correa2016a,Correa2016b,Cruz2017}
\begin{eqnarray}
f(\omega)=\frac{|\mathcal{F}[\omega]|^2}{\int|\mathcal{F}[\omega]|^2d\omega}.
\end{eqnarray}

The CE emerged inspired by information theory, in particular Shannon's theory \cite{Shannon}, so the CE can be described as \cite{Correa2015a,Correa2016a,Correa2016b}
\begin{eqnarray}
S_C[f]=-\sum f_n\ln(f_n),
\end{eqnarray}
which is able to provide us with informational content of settings that are compatible with specific restrictions of a given physical system. So we have some properties that are important to mention, as: $f_n=1/N$ when all $N$ modes carry the same weight. On the other hand, $S_C=0$ if only one mode is present, and we have a discrete DCE maximum when $S_C=\ln N$ \cite{Correa2015a,Correa2016a,Correa2016b}.

Finally, for the case of an arbitrary continuous function $f(\omega)$ on an open interval, the DCE has the form \cite{Gleiser2011, G2,Correa2015a,Correa2015b,Correa2015c,Correa2016a,Correa2016b}
\begin{eqnarray}\label{dce_g}
S_C[f]=-\int \bar{f}(\omega)\ln[\bar{f}(\omega)]d\omega,
\end{eqnarray}
where $\bar{f}(\omega)=f(\omega)/f_{max}(\omega)$ is the normalized modal fraction, and $f_{max}(\omega)$ denotes the maximum fraction.

\subsection{$f_1(T,B)=T+k_1B^{n_1}$}

For the model $f_1(T,B)$, we use the energy density (\ref{q.5}) to find the modal fraction $f(\omega)$. In the case of $n_1=2$, we have that the modal fraction is
\begin{eqnarray}
f(\omega)=\frac{385\pi\omega^2[9\omega^2+32k_1(4-15\omega^2+3\omega^4)]^2}{1152[165+256k_1(600k_1-11)]}\mathrm{csch}^2\Big(\frac{\pi\omega}{2}\Big).
\end{eqnarray}

For $n_1=3$ the modal fraction is
\begin{eqnarray}\nonumber
f(\omega)=\frac{715\pi\omega^2[189\omega^2+32k_1(16420\omega^2-6912-3122\omega^4+95\omega^6)]^2}{8064[19305+14336k_1(1454080k_1+663)]}\mathrm{csch}^2\Big(\frac{\pi\omega}{2}\Big).\\
\end{eqnarray}

\begin{figure}[ht!]
\begin{center}
\begin{tabular}{ccc}
\includegraphics[height=5cm]{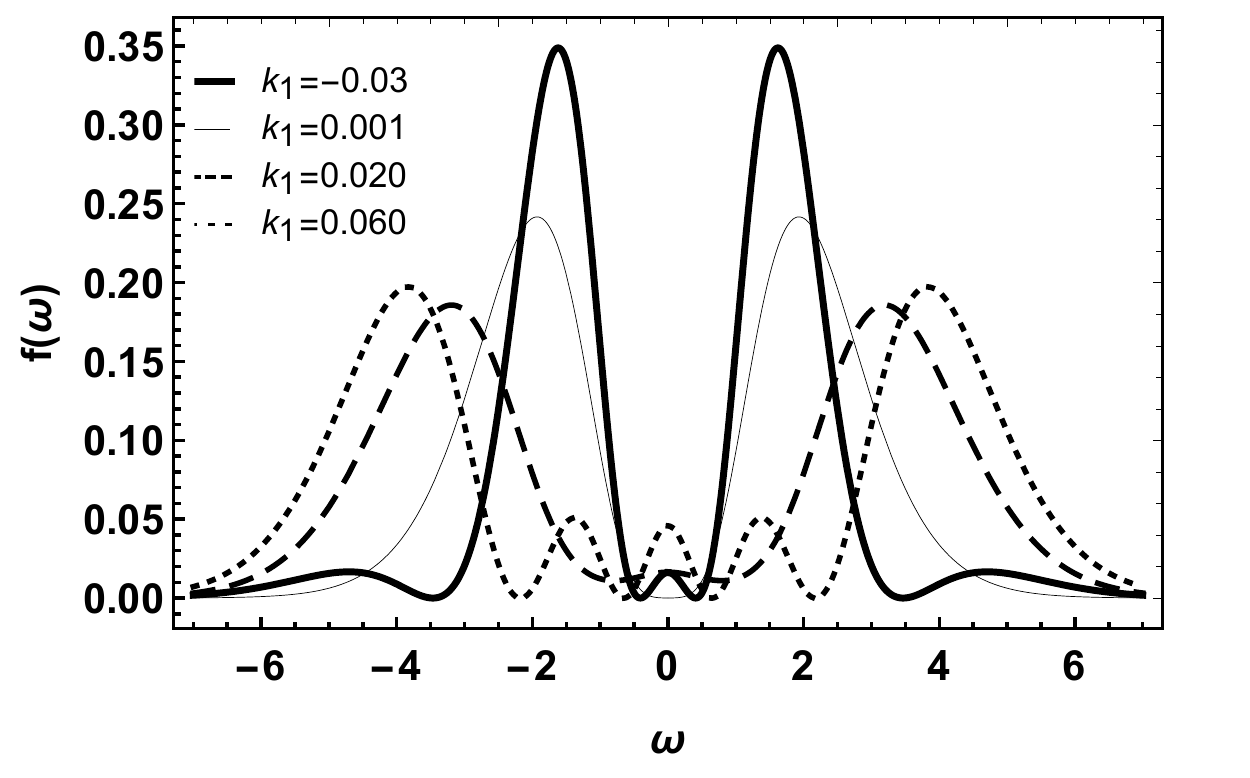}
\includegraphics[height=5cm]{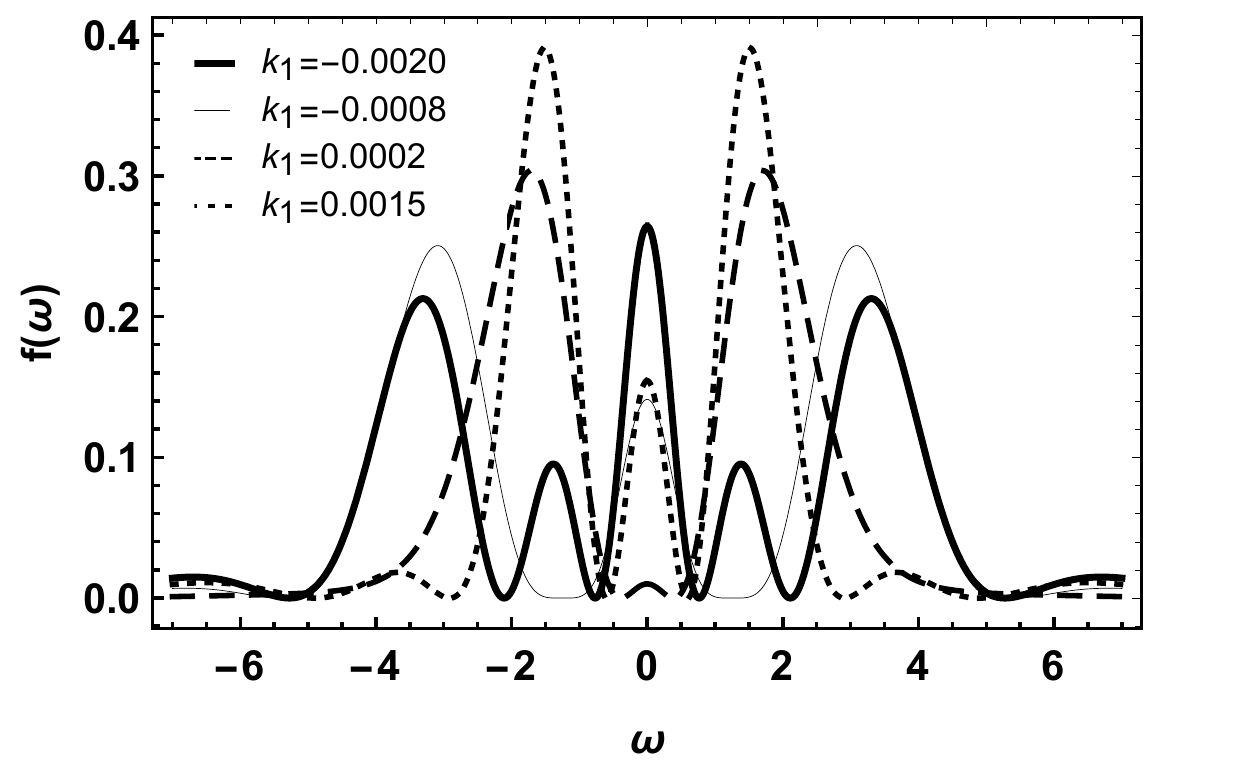}\\ 
(a) \hspace{8 cm}(b)
\end{tabular}
\end{center}
\vspace{-0.7cm}
\caption{Plots of modal fraction for $p=\lambda=1$. (a) $n_1=2$ . (b) $n_1=3$.
\label{fig7}}
\end{figure}

In Fig. \ref{fig7}$a$ for $n_1=2$ and Fig. \ref{fig7}$b$ for $n_1=3$, the plots of modal fractions are shown. The profile of the modal fraction is undergoing modifications as we vary the $n_1$ and $k_1$ parameters. For $n_1=2$, there are two maximal points that are symmetric in relation to origin, and when we increase the value of $k_1$, new symmetric peaks arise. A similar behavior occurs when $n_1=3$. Nonetheless, when $n_1=3$ the modal fraction near the origin becomes more significant.

Let us now connect the discussion in the previous section to the results obtained of the modal fraction. Note that when $k_1$ is positive the modal fraction tends to have several oscillations. However, when $k_1$ takes on negative values as $k=-0.03$ the oscillations tend to decrease so that two symmetric critical points (maximum) and one local maximum in $y=0$ arise. In fact, when comparing the modal fraction with our previous results, we notice that the two peaks of greatest intensity are located near to the points where the scalar field reaches its stability. As a matter of fact, it seems to us that the peak (at $y=0$) indicates the region where the kink starts to transform into a configuration similar to a multi-kink (in this case, similar to double-kink). A consequence of this behavior is the appearance of internal structures in the model indicating the existence of phase transitions and brane splitting. Further, when $n_1=3$ a similar interpretation can be done. Indeed, in this case, we have that the energy density is more intense generating more significant internal structure with a small localization. This leads to smoother one-kink to double-kink phase transitions. Due to the existence of more internal structures close to $y=0$, the modal fraction presents a more intense peak in this region.

\begin{figure}[ht!]
\begin{center}
\begin{tabular}{ccc}
\includegraphics[height=5cm]{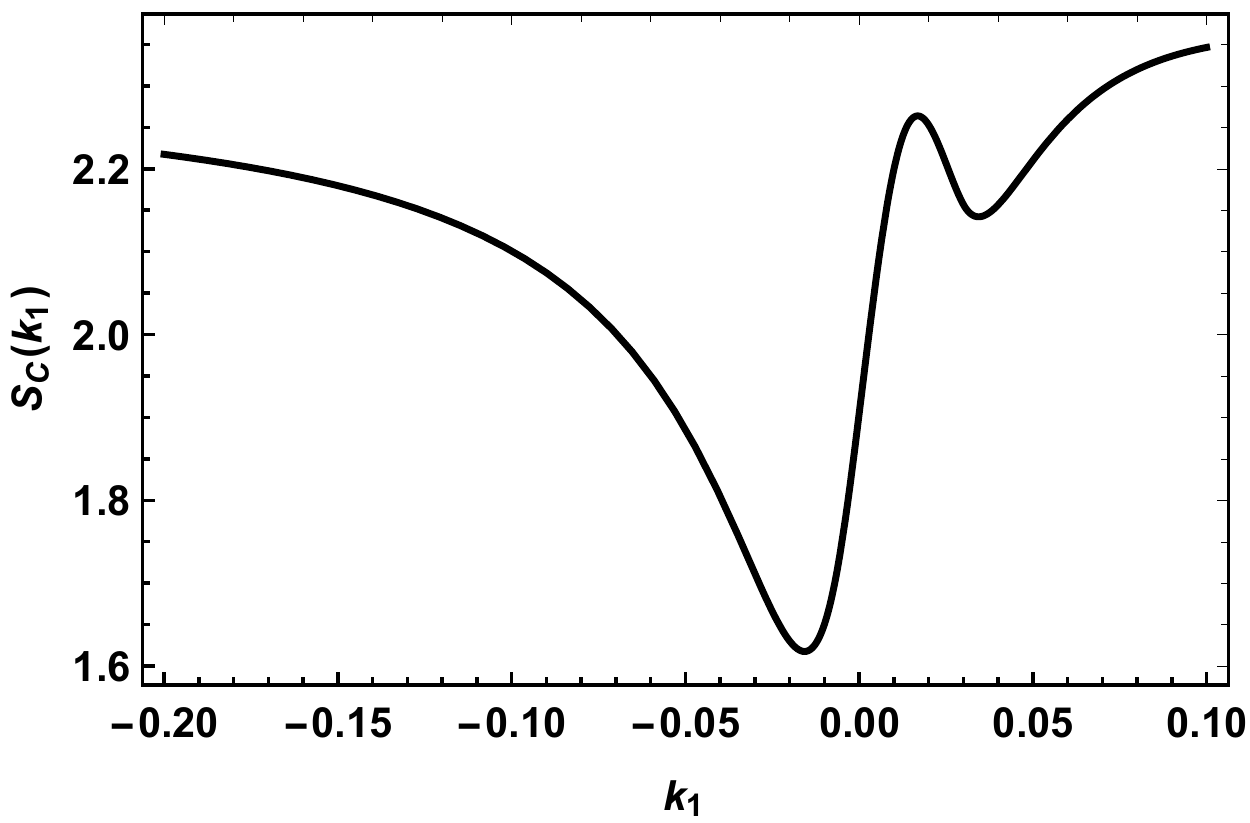}
\includegraphics[height=5cm]{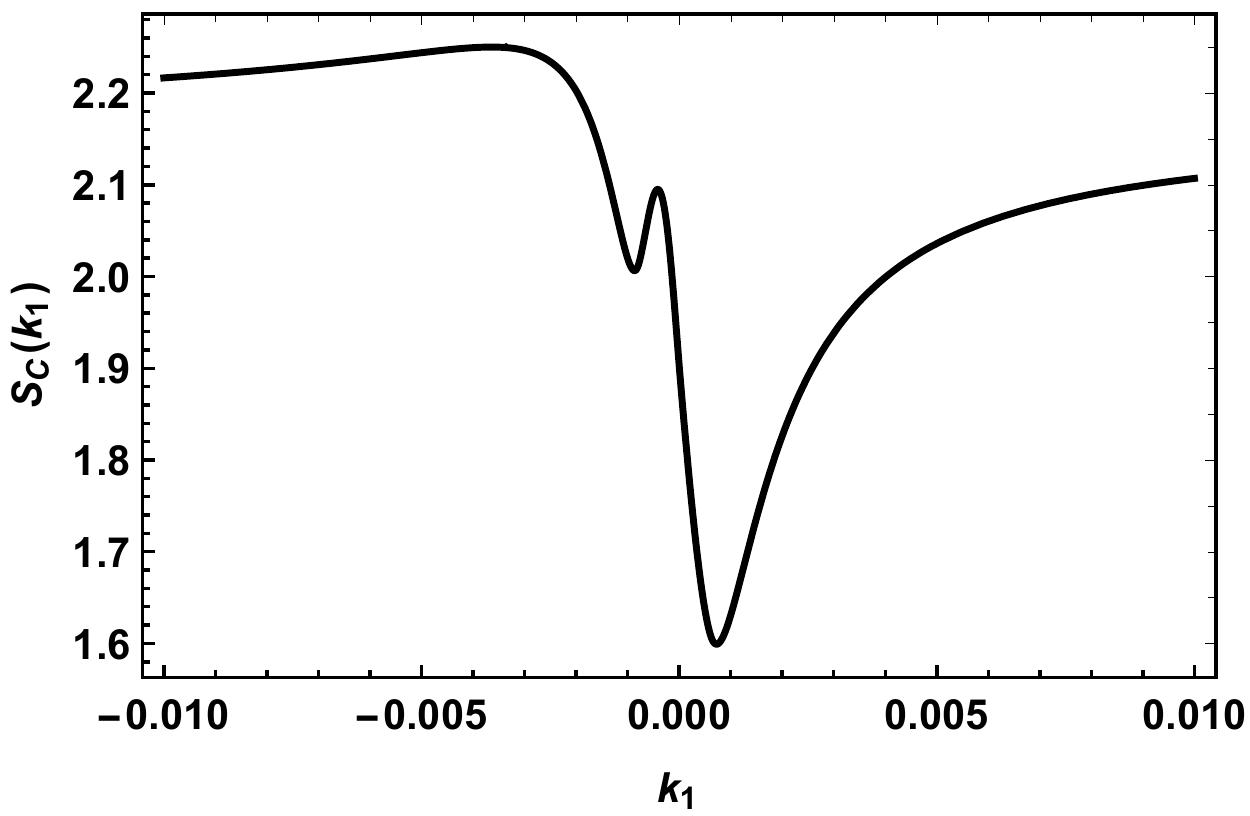}\\ 
(a) \hspace{8 cm}(b)
\end{tabular}
\end{center}
\vspace{-0.7cm}
\caption{Plots of DCE for $p=\lambda=1$. (a) $n_1=2$ . (b) $n_1=3$.
\label{fig8}}
\end{figure}

We stress here that the calculation of DCE is performed by a numerical approach. This occurs because of great complexity to calculate the integral (\ref{dce_g}). The behavior of DCE is shown in Fig. \ref{fig8}. Note that if $n_1=2$ (Fig. \ref{fig8}$a$) the DCE has two minimum points and one maximum point, and the same happens for the case $n_1=3$ (Fig. \ref{fig8}$b$).

In the case $n_1=2$, it is observed that the ``absolute minimum point'' is located in the range $-0.02<k_1<-0.01$. Here it is important to mention that in this range of values occur the emergence of internal structures that are a consequence of the brane split. From topological field theory when $-0.02<k_1<-0.01$ the phase transition, i. e., the transition from kink-like structure to structures that look like double-kink occur. In this range of values the emergence of new domain walls generating the possibility of new topological structures is observed. In turn, for $n_1=3$ with $0<k_1<0.001$, the brane splitting  and the appearance of internal structure in the model are also observed. However, in the case $n_1=3$ for the range of values of $k_1$ due to the DCE profile, it is possible to verify the existence of a structure similar to a triple-kink. This result is interesting and promising as it suggests that more exotic topological structures may appear in the model.

\subsection{$f_2(T,B)=T+k_2(-T+B)^{n_2}$}

For the model $f_2(T,B)$, we use the energy density (\ref{q.6}) to find the modal fraction $f(\omega)$. Initially, when $n_2=2$ the modal fraction is
\begin{eqnarray}
f(\omega)=\frac{77\pi\omega^2[45\omega^2-4k_2(68-30\omega^2+7\omega^4)]^2}{5760[165+8k_2(422k_2-121)]}\mathrm{csch}^2\Big(\frac{\pi\omega}{2}\Big).
\end{eqnarray}

On the other hand, if $n_2=3$ the modal fraction is
\begin{eqnarray}\nonumber
f(\omega)=\frac{1001\pi\omega^2[135\omega^2+2k_2(5128\omega^2-5184-616\omega^4+7\omega^6)]^2}{5760(19305+126048k_2+32905472k_2^2)}\mathrm{csch}^2\Big(\frac{\pi\omega}{2}\Big).\\
\end{eqnarray}

Similar to the $f_1(T,B)$ model, the modal fraction profile in the $f_2(T,B)$ model changes as the parameters $n_2$ and $k_2$ are modified. For $n_2=2$ when the parameter $k_2$ increases the appearance of several oscillations occur,  and arise an absolute maximal point at $y=0$. When $k_2$ is near the range that identifies the appearance of internal structures the modal fraction assumes a new behavior, i. e., the modal fraction has a local minimum that seems to identify the region of phase transition. Further, it is observed the appearance of two symmetrical and absolute maximum points. These points seem to identify the values at which the scalar field reaches stability. It is important to mention that in this transition we have the fields going from kink-like structures to structures that seem to behave like the double-kink when occur the brane splittting. Identical behavior occur when $n_2=3$ and $k_2$ is changed. However, it is interesting to mention that in the second case, due to the existence of multi-kink (or three-kink) solutions, the modal fraction has a more relevant contribution around the origin, as discussed in the previous model.

\begin{figure}[ht!]
\begin{center}
\begin{tabular}{ccc}
\includegraphics[height=5cm]{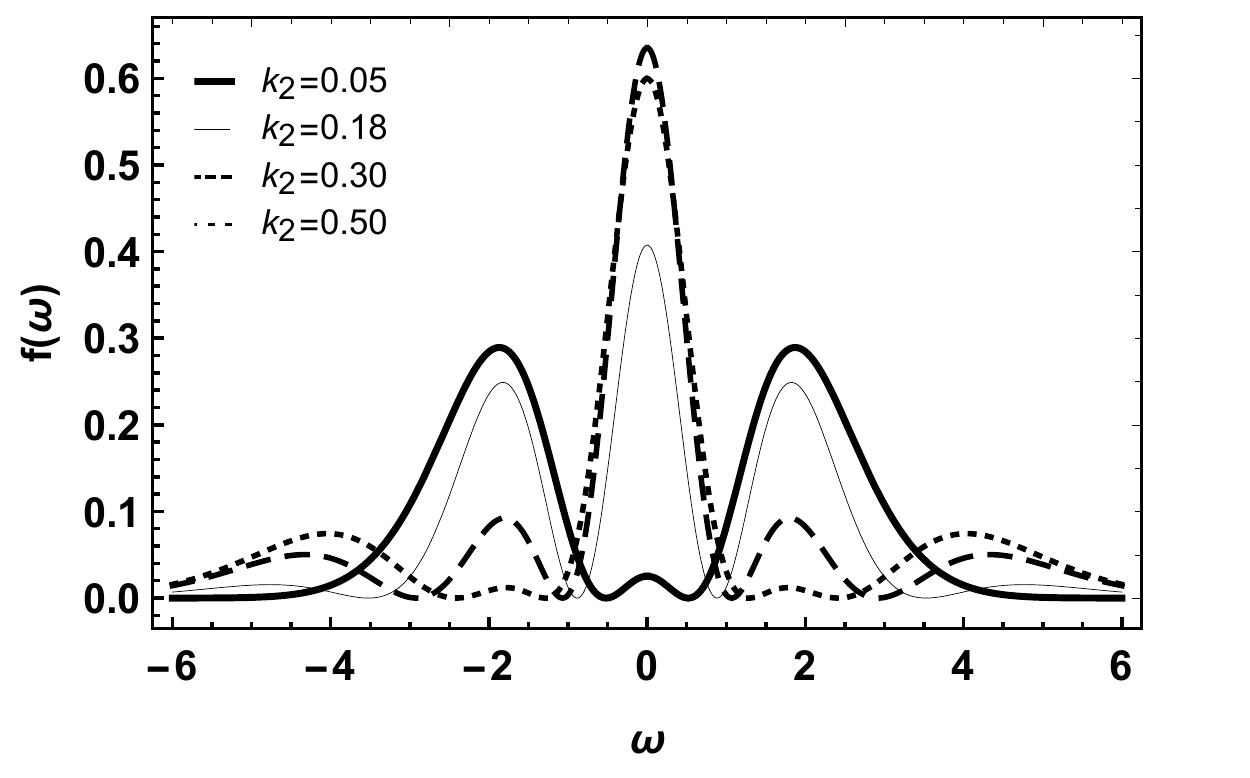}
\includegraphics[height=5cm]{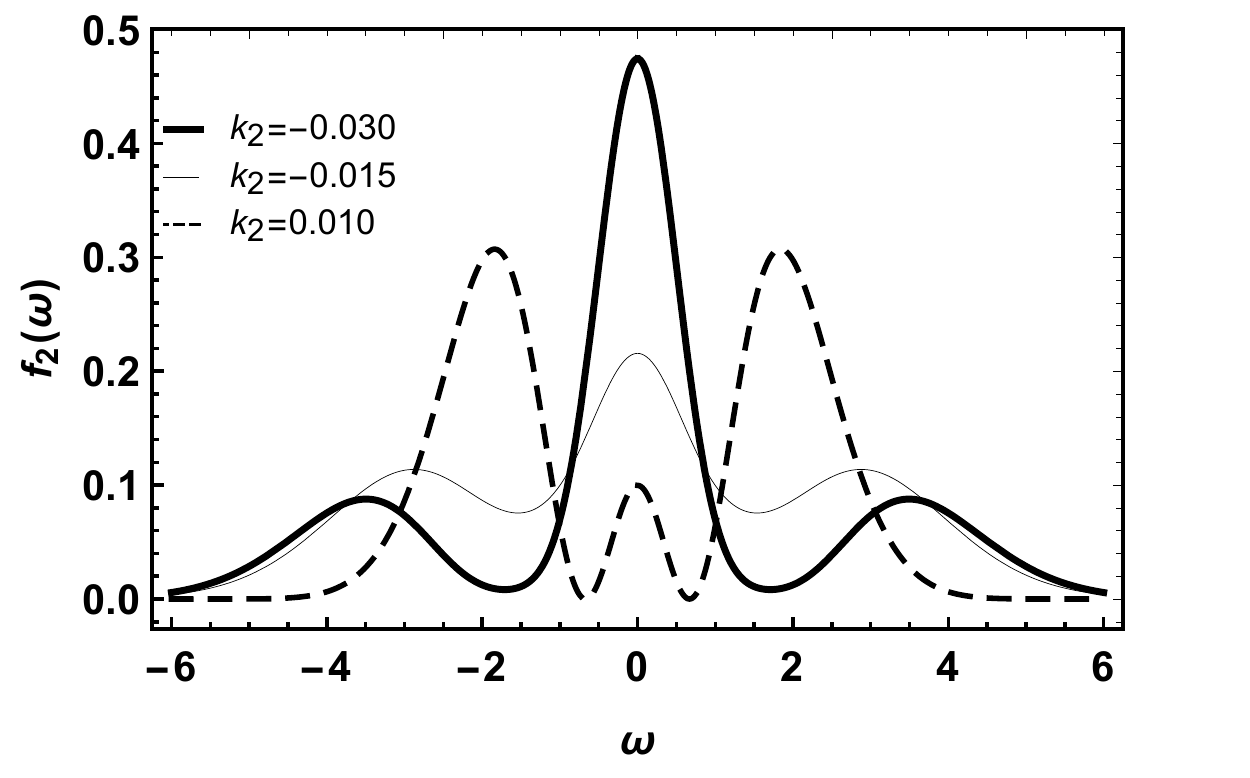}\\ 
(a) \hspace{8 cm}(b)
\end{tabular}
\end{center}
\vspace{-0.7cm}
\caption{Plots of modal fraction for $p=\lambda=1$. (a) $n_2=2$ . (b) $n_2=3$.
\label{fig9}}
\end{figure}

\begin{figure}[ht!]
\begin{center}
\begin{tabular}{ccc}
\includegraphics[height=5cm]{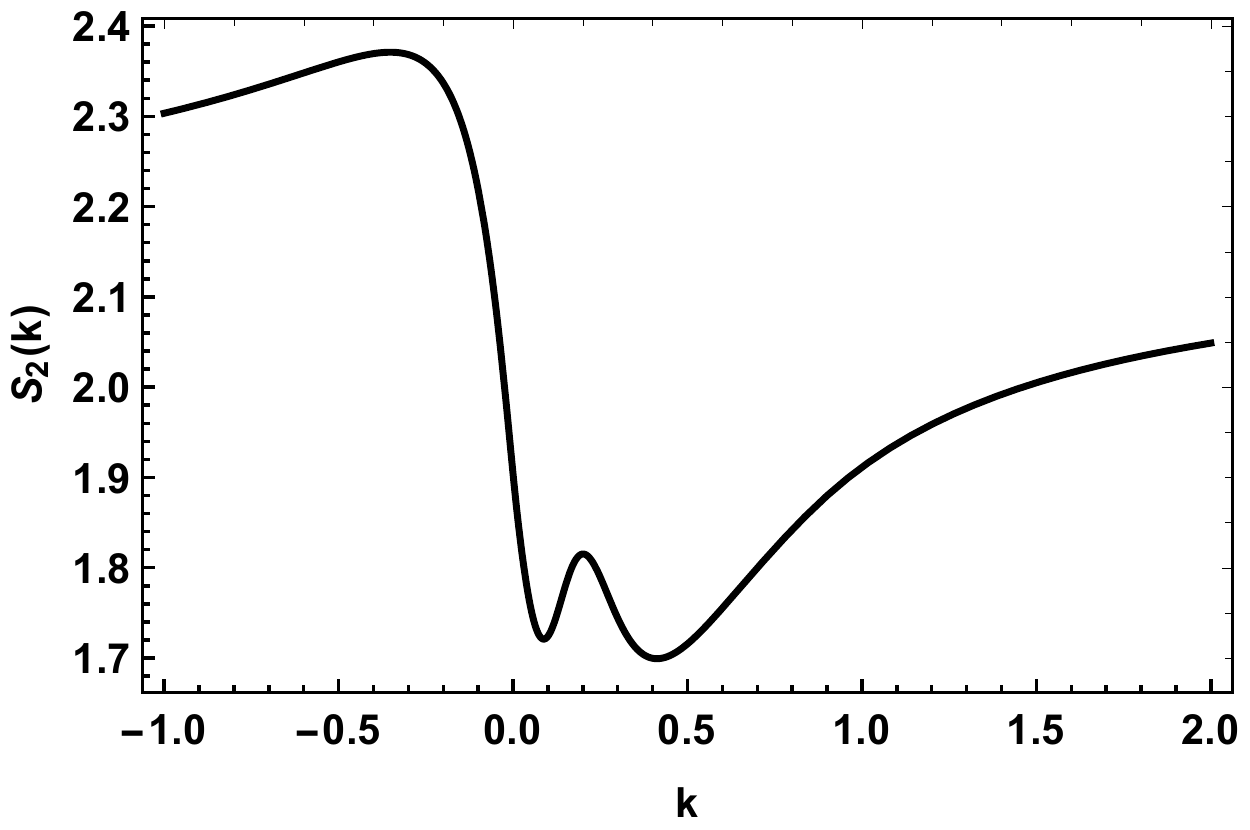}
\includegraphics[height=5cm]{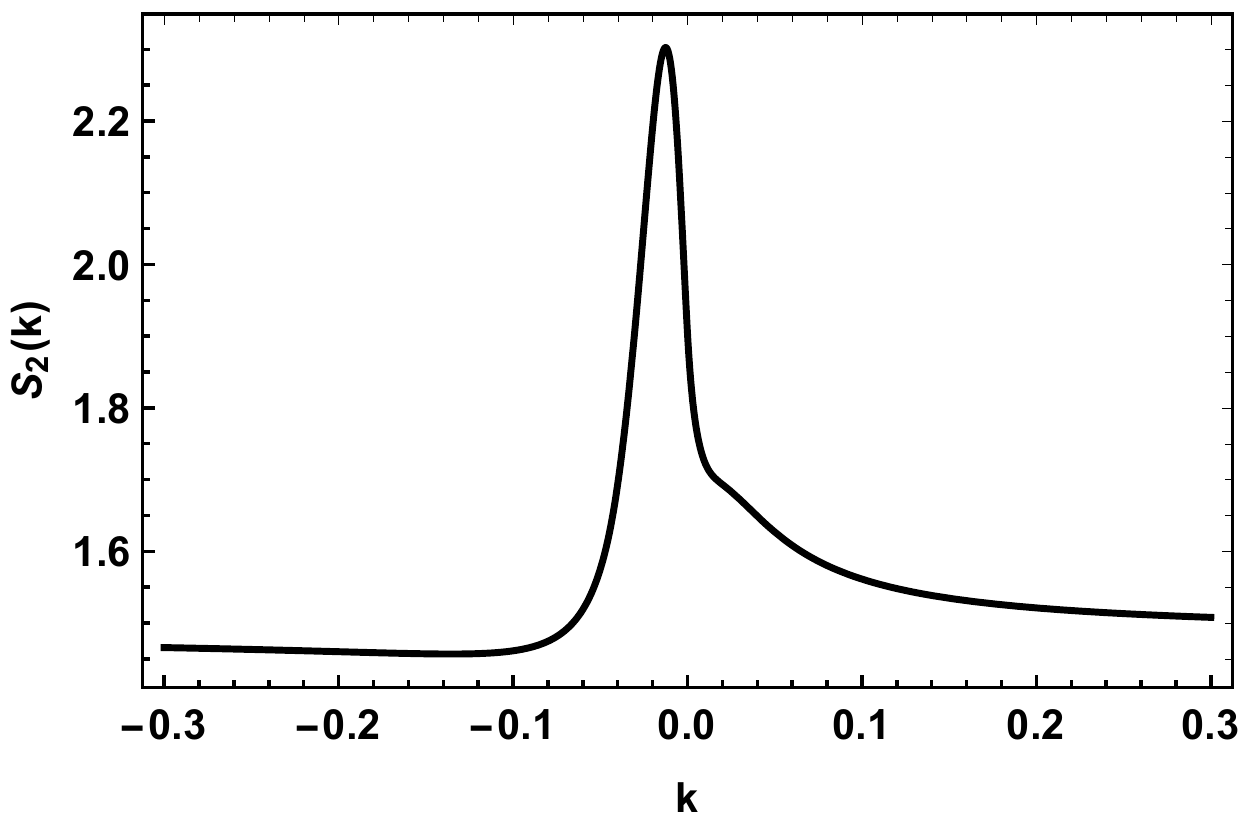}\\ 
(a) \hspace{8 cm}(b)
\end{tabular}
\end{center}
\vspace{-0.7cm}
\caption{Plots of DCE for $p=\lambda=1$. (a) $n_2=2$ . (b) $n_2=3$. 
\label{fig10}}
\end{figure}

As in the previous case, to obtain the DCE result of the model, the calculations were numerically performed. The behavior of DCE is shown in Fig. \ref{fig10}. Note that for $n_2=2$ (Fig. \ref{fig9}$a$) the DCE presents something quite different, where the minimum two points are very close. Further, in the case $n_2=3$ (Fig. \ref{fig9}$b$), the DCE have an absolute maximal point followed by an almost exponential decay. 

In the case $n_2=2$, it is observed that the first minimal is located in the range $0<k_2<0.1$. For this range of values also occurs the emergence of internal structures that are a consequence of the brane splitting. For $0<k_2<0.1$, the phase transition occurs, i. e., the transition from kink-like structure to structures that look like double-kink. Something unusual happens for $n_2=3$. The DCE maximal point located at $-0.01<k_2<0.12$, marks the division of the brane (Fig. \ref{fig4}$b$) and the phase transition of the scalar field (Fig. \ref{fig6}). In other words, the critical point in DCE seems to identify the location of multiple walls that generate multi-kink configurations, i. e., it seems to us that this region registers multiple phase transitions in theory.

\section{Final remarks}
\label{sec3}

In this work, we studied teleparallel gravity models of the type $f_1 (T,B)=T+k_1 B^{n_1}$ and $f_2(T,B)=-T+k_{2}(-T+B)^{n_2}$. In both cases, it was observed that the parameters $k_{1,2}$ and $n_{1,2}$ are responsible for the division of the brane that generates internal structures that are related to the appearance of new domain walls. In fact, with the appearance of these walls, new classes of topological structures for the matter field were obtained.

Taking into account the criteria of the approach presented by Gleiser and Stamatopoulos \cite{Gleiser2011}, the DCE is directly related to the energy of the system, where the smaller (larger) the DCE, the smaller (greater) the energy of the solutions. DCE is able to provide criteria to control the stability of settings based on the informative content of your profiles \cite{G2}. Thus, with the help of DCE it was possible to select the most prominent solutions for our $f_{1,2}(T,B)$ models, analyzing some specific values of the $n_{1,2}$ and $k_{1,2}$ parameters.

In the model $f_1(T,B)$ with $n_1=2$, the DCE minimum point located around $-0.02<k_1<-0.01$ marks the phase transition of the brane, which tends to splitting. The same is true for $n_1=3$, with the DCE minimal point located around $0<k_1<0.001$. On the other hand, for $f_2(T,B)$ with $n_2=2$, the DCE minimum point located around $0<k_2<0.1$ marks the phase transition of the brane, which tends to splitting. When $n_2=3$, the brane phase transition is marked by the maximum point of the DCE which is located in the interval $-0.01<k_2<0.12$. Here it is important to mention that the topological configurations most likely to be found in the system are described by the range of values of the parameters $k_{1,2}$ mentioned above.

Finally,  it is observed that DCE provides a complementary perspective to investigate the phase transition, splitting of the brane, and the arising of new class of structures in braneworld in the context of modified teleparallel gravity theories, or other theories of gravitation.

\section*{Acknowledgments}
\hspace{0.5cm} The authors thank the Conselho Nacional de Desenvolvimento Cient\'{\i}fico e Tecnol\'{o}gico (CNPq), n$\textsuperscript{\underline{\scriptsize o}}$ 309553/2021-0 (CASA), and Coordenaçao de Aperfeiçoamento do Pessoal de Nível Superior (CAPES), for financial support.


\begin{thebibliography}{99}

\bibitem{WMAP}
WMAP collaboration, C. Bennett et al., Astrophys. J. Suppl. {\bf 148}, 1 (2003) [astro-ph/0302207]; 

\bibitem{WMAP1}
WMAP collaboration, D.N. Spergel et al., Astrophys. J. Suppl. {\bf 148}, 175 (2003) [astro-ph/0302209].

\bibitem{Supernova}
Supernova Cosmology Project collaboration, S. Perlmutter et al., Nature {\bf 391}, 51 (1998).

\bibitem{Supernova1}
Supernova Cosmology Project collaboration, S. Perlmutter et al.,  Astrophys. J. {\bf 517}, 565 (1999).

\bibitem{SDSS}
SDSS collaboration, D.J. Eisenstein et al., Astrophys. J. {\bf 633}, 560 (2005).

\bibitem{Sharif}
M. Sharif, and M. Zubair, J. Cosm. Astrophys. \textbf{2012}, 28 (2012).

\bibitem{Sahni}
V. Sahni, and A. A. Starobinsky, Int. J. Mod.
Phys. D {\bf 9} 373 (2000).

\bibitem{Sahni1}
V. Sahni, Lect. Notes Phys. {\bf 653}, 141 (2004).

\bibitem{Padmana}
T. Padmanabhan, Gen. Relativ. Gravit. {\bf 40}, 529 (2008).

\bibitem{Caldwell}
R. Caldwell,  Phys. Lett. B {\bf 545}, 23 (2002).

\bibitem{Nojiri}
S. Nojiri, and S.D. Odintsov, Phys. Lett. B {\bf 562}, 147  (2003).

\bibitem{Feng}
B. Feng, X.-L. Wang and X.-M. Zhang, Phys. Lett. B {\bf 607}, 35 (2005).

\bibitem{Kamenshchik}
A. Y. Kamenshchik, U. Moschella, and V. Pasquier, Phys. Lett. B {\bf 511}, 265 (2001).

\bibitem{Sotiriou}
T.P. Sotiriou and S. Liberati, Ann. Phys. {\bf 322}, 935 (2007).

\bibitem{Sotiriou1}
T.P. Sotiriou and V. Faraoni, Rev. Mod. Phys. {\bf 82}, 451 (2010).

\bibitem{Kaluza}
T. Kaluza, Math. Phys. \textbf{1921}, 966 (1921).

\bibitem{Klein}
O. Klein, Nature \textbf{118}, 516 (1926).

\bibitem{Wang}
J. Wang, Wen-Di Guo, Zi-Chao Lin, and Yu-Xiao Liu, Phys. Rev. D \textbf{98}, 084046 (2018). 

\bibitem{Arkani}
N. Arkani-Hamed, S. Dimopoulos, and G. R. Dvali, Phys. Lett. B {\bf 429}, 263 (1998).

\bibitem{Randall&Sundrum}
L. Randall, and R. Sundrum, Phys. Rev. Lett. {\bf 83}, 3370 (1999).

\bibitem{Randall&Sundrum1}
L. Randall, and R. Sundrum, Phys. Rev. Lett. \textbf{83}, 4690 (1999). 

\bibitem{Bogdanos}
C. Bogdanos, A. Dimitriadis, and K. Tamvakis, Phys. Rev. D
{\bf 74}, 045003 (2006).

\bibitem{Bogdanos1}
C. Bogdanos, J. Phys. Conf. Ser. {\bf 68}, 012045 (2007).

\bibitem{Yang}
K. Yang, Y.-X. Liu, Y. Zhong, X.-L. Du, and S.-W. Wei, Phys. Rev. D {\bf 86}, 127502 (2012).

\bibitem{Farakos}
K. Farakos, G. Koutsoumbas, and P. Pasipoularides, Phys. Rev. D {\bf 76}, 064025 (2007).

\bibitem{Liu}
Y.-X.Liu, F.-W.Chen, H.Guo,and X.-N.Zhou, J. High Energy Phys. {\bf 05}, (2012) 108.

\bibitem{Guo}
H. Guo, Y.-X. Liu, Z.-H. Zhao, and F.-W. Chen, Phys.
Rev. D {\bf 85}, 124033 (2012).

\bibitem{Parry}
M. Parry, S. Pichler, and D. Deeg, J. Cosmol. Astropart. Phys. {\bf 04} (2005) 014.

\bibitem{Petrov}
D. Bazeia, A. S. Lobao, R. Menezes, A. Yu. Petrov, and A. J. Silva, Phys, Lett. B \textbf{729}, 127 (2014).

\bibitem{Petrov1}
V. I. Afonso, D. Bazeia, R. Menezes, and A. Yu. Petrov, Phys. Lett. B \textbf{658}, 71 (2007).

\bibitem{Bazeia}
D. Bazeia, L. Losano, R. Menezes, G. J. Olmo, and D.
Rubiera-Garcia, Eur. Phys. J. C {\bf 75}, 569 (2015).

\bibitem{Gu}
B.-M. Gu, B. Guo, H. Yu, and Y.-X. Liu, Phys. Rev. D {\bf 92}, 024011 (2015).

\bibitem{Olmo}
C. Barragán, and G. J. Olmo. Phys. Rev. D {\bf 82},  084015 (2010).

\bibitem{Olmo1}
G. J. Olmo, Intern. J. Mod. Phys. D \textbf{20}, 413 (2011).

\bibitem{Cai}
Y.-F. Cai, S. Capozziello, M. De Laurentis, and E. N.
Saridakis, Rep. Prog. Phys. {\bf 79}, 106901 (2016).

\bibitem{Behdoodi}
A. Behboodi, S. Akhshabi, and K. Nozari, Phys. Lett. B {\bf 723}, 201 (2013).

\bibitem{Nozari}
K. Nozari, and N. Sadeghnezhad, Intern. J. Geom. Meth. Mod. Phys. {\bf 16}, 1950042 (2019).

\bibitem{Moreira2021b}
A. R. P. Moreira, J. E. G. Silva, and C. A. S. Almeida, Europ. Phys. J. C {\bf 81}, 1-9 (2021).

\bibitem{Moreira2021a}
A. R. P. Moreira, J. E. G. Silva, F. C. E. Lima, and C. A. S. Almeida, Phys. Rev. D \textbf{103}, 064046 (2021). 

\bibitem{Kofinas}
G. Kofinas, and E. N. Saridakis, Phys. Rev. D {\bf 90}, 084044 (2014).

\bibitem{Kofinas1}
G. Kofinas, and E. N. Saridakis, Phys. Rev. D {\bf 90}, 084045 (2014).

\bibitem{Chatto}
S. Chattopadhyay, A. Jawad, D. Momeni, and R. Myrzakulov, Astrophys. Space Sci. {\bf 353}, 279 (2014).

\bibitem{Gomez}
D. Saez-Gomez, C. S. Carvalho, F. S. N. Lobo and I. Tereno, Phys. Rev. D {\bf 94}, 024034 (2016).

\bibitem{Harko}
T. Harko, F. S. N. Lobo, G. Otalora, and E. N. Saridakis, J. Cosm. Astrop. Phys. \textbf{2014}, 21 (2014).

\bibitem{Bahamonde}
S. Bahamonde, and S. Capozziello, Eur. Phys. J. C {\bf 77}, 107 (2017).

\bibitem{Franco}
G. A. R. Franco, C. Escamilla-Rivera and J. Levi Said, Eur. Phys. J. C {\bf 80}, 677 (2020).

\bibitem{Rivera}
C. Escamilla-Rivera, and J. Levi Said, Class. Quant. Grav. {\bf 37}, 165002 (2020).

\bibitem{Said}
S. Bahamonde,  K. F. Dialektopoulos, C. Escamilla-Rivera, G. Farrugia, V. Gakis, M. Hendry, M. Hohmann, J. L. Said, J.  Mifsud, and E.  Di Valentino, {\it Teleparallel Gravity: From Theory to Cosmology}. arXiv 2021, arXiv:2106.13793v3.

\bibitem{BLi2010}
B.~Li, T.~P.~Sotiriou and J.~D.~Barrow,
Phys. Rev. D \textbf{83} (2011), 064035.

\bibitem{Sotiriou2010}
T.~P.~Sotiriou, B.~Li and J.~D.~Barrow,
Phys. Rev. D \textbf{83} (2011), 104030.

\bibitem{Capozziello2019}
S.~Capozziello, M.~Capriolo and L.~Caso,
Eur. Phys. J. C \textbf{80} (2020) no.2, 156

\bibitem{Bahamonde1}
S. Bahamonde, V. Gakis, S. Kiorpelidi, T. Koivisto, J. L. Said, and E. N. Saridakis, Eur. Phys. J. C {\bf 81}, 53 (2021).

\bibitem{Pourbagher}
A. Pourbagher, and A. Amani, Astrophysics Space and Science, {\bf 364}, 1-8 (2018).

\bibitem{Pourbagher1}
A. Pourbagher, and A. Amani, Modern Phys. Lett. A \textbf{35}, 2050166 (2020).

\bibitem{Abedi}
H. Abedi and S. Capozziello, Eur. Phys. J. C {\bf 78}, 474 (2018).

\bibitem{Bhattacharjee}
S. Bhattacharjee, Phys. Dark Universe {\bf 30}, 100612 (2020).

\bibitem{GleiserSowinki}
M. Gleiser, and D. Sowinski, Phys. Rev. D \textbf{98}, (2018) 056026.

\bibitem{Ranada}
A. F. Ra\~{n}ada, Lett. Math. Phys. {\bf 18}, 97 (1989).

\bibitem{Dva}
G. Dvali, I. I. Kogan, and M. Shifman. Phys. Rev. D {\bf 62}, 106001 (2000).

\bibitem{LPA}
F. C. E. Lima, A. Yu. Petrov, and C. A. S. Almeida, Phys. Rev. D \textbf{103}, 096019 (2021).

\bibitem{GS}
M. Gleiser, and N. Stamatopoulos, Phys. Lett. B \textbf{713} (2012) 304.

\bibitem{Correa2015a}
R.~A.~C.~Correa, P.~H.~R.~S.~Moraes, A.~de Souza Dutra and R.~da Rocha, Phys. Rev. D \textbf{92} (2015) no.12, 126005.

\bibitem{Gleiser2011}
M.~Gleiser and N.~Stamatopoulos, Phys. Lett. B \textbf{713} (2012), 304-307.

\bibitem{G1}
M. Gleiser and N. Stamatopoulos, Phys. Rev. D {\bf 86}, 045004 (2012).

\bibitem{G2}
M. Gleiser and D. Sowinski, Phys. Lett. B {\bf 727}, 272 (2013).

\bibitem{G3}
M. Gleiser and N. Graham, Phys. Rev. D {\bf 89}, 083502 (2014).

\bibitem{Correa2015c}
R.~A.~C.~Correa and R.~da Rocha,
Eur. Phys. J. C \textbf{75} (2015) no.11, 522.

\bibitem{Correa2016a}
R.~A.~C.~Correa, D.~Monteiro Dantas, P.~H.~R.~da Silva Moraes, A.~de Souza Dutra and C.~A.~S.~de Almeida,
Annalen Phys. \textbf{530} (2018) no.7, 1700188.

\bibitem{Correa2016b}
R.~A.~C.~Correa, P.~H.~R.~S.~Moraes, A.~de Souza Dutra, W.~de Paula and T.~Frederico,
Phys. Rev. D \textbf{94} (2016) no.8, 083509.

\bibitem{Correa2016p}
R.~A.~C.~Correa, D.~M.~Dantas, C.~A.~S.~Almeida and R.~da Rocha,
Phys. Lett. B \textbf{755} (2016), 358-362.

\bibitem{Cruz2017}
W.~T.~Cruz, D.~M.~Dantas, R.~A.~C.~Correa and C.~A.~S.~Almeida,
Phys. Lett. B \textbf{772} (2017), 592-598.

\bibitem{Cruz2018}
W.~T.~Cruz, D.~M.~Dantas, R.~V.~Maluf and C.~A.~S.~Almeida,
Annalen Phys. \textbf{531} (2019) no.10, 1900178.

\bibitem{Correa2015b}
R.~A.~C.~Correa and P.~H.~R.~S.~Moraes,
Eur. Phys. J. C \textbf{76} (2016) no.2, 100.

\bibitem{Aldrovandi}
R.~Aldrovandi and J.~G.~Pereira,
``Teleparallel Gravity: An Introduction,''
(Springer,Berlin, 2013).

\bibitem{Abedi2017}
H.~Abedi and S.~Capozziello,
Eur. Phys. J. C \textbf{78}, 474 (2018).

\bibitem{Yang2012}
J.Yang, Y.-L.~Li, Y.~Zhong and Y.~Li,
 Phys.\ Rev.\ D {\bf 85}, 084033 (2012).

\bibitem{Gremm1999}
M.~Gremm,
Phys. Lett. B \textbf{478}, 434 (2000).

\bibitem{D}
D. Bazeia, D. A. Ferreira, and M. A. Marques, Europ. Phys. J. C {\bf 81}, 619 (2021).

\bibitem{Hind}
M. Hindmarsh, K. Rummukainen, and D.J. Weir, Phys. Rev. Lett. {\bf 117}, 251601 (2016).

\bibitem{Hart}
B. Hartmann, F. Michel, and P. Peter, Phys. Rev. D {\bf 96}, 123531 (2017).

\bibitem{Cao}
X.-F. Cao and Y.-W. Yu, Phys. Rev. D {\bf 97}, 023022 (2018).

\bibitem{Shannon}
C. E. Shannon, The Bell system technical journal {\bf 27}, 379 (1948).


\end{thebibliography}

\end{document}